%% file: ms.tex
\documentclass[journal]{IEEEtran}

\usepackage{amsmath,amsfonts,amssymb,amsthm}
\usepackage[inline]{enumitem}
\usepackage{multirow,hhline,graphicx,tabularx,makecell}
\usepackage{subfig}
\usepackage{setspace}
\usepackage{booktabs,longtable,array,wrapfig,float,colortbl,pdflscape,tabu}
\usepackage{threeparttable,threeparttablex,makecell,xcolor}
\usepackage{cite}
\usepackage{tikz}
\usepackage{lscape,rotating}
\usepackage[normalem]{ulem}
\usepackage{epstopdf}
\usepackage{hyperref}
\usepackage[multiple]{footmisc}

\newtheorem{hypothesis}{Hypothesis}
\newtheorem{postulate}{Postulate}
\renewcommand{\Vec}[1]{\mathbf{#1}}

\hypersetup{
    citecolor=blue,
    colorlinks=true,
    linkcolor=purple,    
    urlcolor=blue
}

\begin{document}
\title{A Review \& Framework for Modeling Complex Engineered System Development Processes}

\author{John~Meluso, Jesse~Austin-Breneman, James P.~Bagrow, \& Laurent H\'{e}bert-Dufresne \thanks{This work has been submitted to the IEEE for possible publication. Copyright may be transferred without notice, after which this version may no longer be accessible. Manuscript received [to be completed by editor]. The work of J. Meluso, J. Bagrow, \& L. H\'{e}bert-Dufresne was supported in part by Google Open Source under the Open-Source Complex Ecosystems And Networks (OCEAN) project.}
\thanks{\textbf{John Meluso} is the OCEAN Postdoctoral Fellow at the Vermont Complex Systems Center at the University of Vermont, Burlington, VT 05405.}
\thanks{\textbf{Jesse Austin-Breneman} is Assistant Professor of Mechanical Engineering at the University of Michigan, Ann Arbor, MI 48109.}
\thanks{\textbf{James Bagrow} is Associate Professor of Mathematics \& Statistics at the University of Vermont, Burlington, VT 05405.}
\thanks{\textbf{Laurent H\'{e}bert-Dufresne} is Assistant Professor of Computer Science at the University of Vermont, Burlington, VT 05405.}}

\maketitle

\begin{abstract}
    Developing complex engineered systems (CES) poses significant challenges for engineers, managers, designers, and businesspeople alike due to the inherent complexity of the systems and contexts involved. Furthermore, experts have expressed great interest in filling the gap in theory about how CES develop. This article begins to address that gap in two ways. First, it reviews the numerous definitions of CES along with existing theory and methods on CES development processes. Then, it proposes the ComplEx System Integrated Utilities Model (CESIUM), a novel framework for exploring how numerous system and development process characteristics may affect the performance of CES. CESIUM creates simulated representations of a system architecture, the corresponding engineering organization, and the new product development process through which the organization designs the system. It does so by representing the system as a network of interdependent artifacts designed by agents. Agents iteratively design their artifacts through optimization and share information with other agents, thereby advancing the CES toward a solution. This paper describes the model, conducts a sensitivity analysis, provides validation, and suggests directions for future study.
\end{abstract}


\section{Introduction}
\label{sec:introduction}

Society increasingly relies on healthcare systems, stock markets, automotive systems, national defense programs, and countless other designed systems; however, these systems continually grow more difficult to develop and manage due to the complexity within and around them. These \textit{complex engineered systems} (CES) are large sets of highly-interacting engineered artifacts with a defined purpose.\footnote{An \textit{artifact} is any piece of technology designed to serve a specific purpose \cite{DeWeck2011}, often used as ``an umbrella term for any technical product of human minds including physical parts, software, processes, information, etc.'' \cite{Meluso2019}.} Often ``difficult to describe, understand, predict, manage, design, or change''  \cite{DeWeck2011} they are composed of many heterogeneous elements, characterized by nonlinear interactions at multiple levels of organization and abstraction, and exhibit complex behaviors which emerge from those interactions \cite{Braha2006}.

To date, government and industry organizations alike manage systems via systems engineering techniques \cite{Kapurch2007}, Deming's quality control methods \cite{Deming1986}, and six sigma principles \cite{Schroeder2008}. However, traditional methods have not kept pace with the varying scales and increasing interactions between elements of systems. The cost and schedules of engineering projects have grown exponentially \cite{Augustine1997, Collopy2012a, Maddox2013}, leading to frequent overruns \cite{Bloebaum2012, Bloebaum2012a}, billions of dollars in losses, and even in lives \cite{Rogers1986, Bar-Yam2003, GehmanJr2003}. Such pervasive failures mean that rather than ``exceptional,'' failures have instead become ``normal'' or expected \cite{Perrow2011}. Made more challenging, organizational factors often create the adverse conditions in which society feels the effects of technical and ``human'' errors \cite{Madni2009}.

While traditional systems engineering methods still hold value, leading government, industry, and academic voices have expressed a ``dire need'' \cite{Collopy2015} for deeper theoretical understandings of CES, their development processes, and the organizations that create them \cite{Minai2006, Bloebaum2012, Bloebaum2012a, Collopy2012a, Triantis2014, Collopy2015, Salvucci-favier2016}. Throughout his work on behalf of the United States National Science Foundation (NSF), the United States Department of Defense (DoD), and the International Council on Systems Engineering (INCOSE), Collopy \& colleauges identified several areas ``ripe for exploration'' to address these needs including abstraction and model-based systems design \cite{Collopy2015a, Topcu2020}. He suggests that ``theory could make a real difference'' toward understanding CES development processes (CESDPs) \cite{Collopy2015}.

Researchers have begun to build theory for traditional systems engineering processes \cite{Safarkhani2020} and frameworks for particular CES contexts such as autonomous vehicles \cite{Young2017} or Internet of Things devices \cite{Cecil2019, Fortino2021}. Nonetheless, few efforts have suggested frameworks through which to build theoretical understandings of the complexity involved in CES more broadly and their requisite development processes \cite{Collopy2015a}. Increasingly, scholars recommend utilizing \textit{systems theory} to assess such problems, a set of concepts rooted in complex systems and management scholarship \cite{Adams2014, Keating2014, Poole2014, Keating2020}, acknowledging the need for agent-based methods to explore relative independence, discrete events, and interactive dynamics \cite{Katina2020}.

Fortunately, recent advances make such analysis possible. Network theory facilitates simulated representations of CES architectures \cite{Barabasi1999a, Clauset2009, Sosa2011}, and agent-based modeling grants access to macro-level outcomes of micro-level decisions and interactions \cite{Schelling1971, Axelrod1997, Smith2017}. Systems engineering research has explored alternative system formulations such as Value-Driven Design \cite{Collopy2011, Collopy2012} while engineering design research has formulated Multidisciplinary Design Optimization \cite{Martins2013}, each of which lends itself to abstract representations of CES development. Likewise, organizational research is beginning to explore computational methods of representing organizational processes \cite{Carley2002, Hong2004, Harrison2007, Fioretti2012, Morgan2015, Clement2018} as engineering follows suit \cite{Crowder2012, McComb2015, McComb2016, McComb2017, Meluso2018, Meluso2019}.

In light of these advances, this paper builds on the previous work of Meluso et al. \cite{Meluso2019} to propose a modeling framework for analyzing how organizations develop CES. The framework combines techniques from systems engineering, complex systems, engineering design, and organizational theory to form the ComplEx System Integrated Utilities Model (CESIUM).

The rest of this article is organized as follows. First, it summarizes the literature defining CES, the constituent elements of CESDPs, and the methods for scientifically studying CESDPs (Sec. \ref{sec:background}). It then describes the model and simulation methods of CESIUM as an abstract, but grounded, representation of a CESDP (Sec. \ref{sec:description}), before characterizing the model through parameter sweeps, discussing validation, and exemplifying its potential for theory formation (Sec. \ref{sec:analysis}). The paper closes by generalizing the framework to facilitate the development of a more robust theory of CES development.

\section{Background}
\label{sec:background}

Several disciplines touch, and uniquely define, the complex systems constructed by people. As such, this section starts by delineating the definitions of CES (Sec. \ref{subsec:background_definitions}). From this common understanding, it describes the constituent elements of CESDPs (Sec. \ref{subsec:background_constituents}). Then, it describes the methods through which researchers study CESDPs (Sec. \ref{subsec:background_methods}) before detailing the literature specific to the methods of CESIUM (Sec. \ref{subsec:background_influences}).

\subsection{Definitions of Complex Engineered Systems}
\label{subsec:background_definitions}


Scholars continue to debate what to call the systems of interacting artifacts that people design, and with good reason. Different definitions emphasize different system characteristics, the relationships between the artifacts within the system, and the relationship between the system and the outside world. At the least, it creates something of a ``branding problem'' when university departments adopt more than 20 unique names to describe CES \cite{DeWeck2011}. A few particular terms are worth highlighting, though, as they reveal the core elements necessary to describe CESDPs.

\subsubsection{Complex Systems}
\label{subsubsec:background_definitions_complexsystems}

Conceptually, complex systems appear in disciplines from biology \cite{Jacob1977, Kitano2004} to physics \cite{Barabasi1999a} to sociology \cite{Watts1998, Bruch2013}. The interdisciplinary field of complex systems grew out of these and several other disciplines in the latter half of the twentieth century with a few core methods for understanding adaptation, information, and collective behaviors \cite{Mitchell2009}. Newman \cite{Newman2011} notes that ``there is no precise technical definition of a `complex system,' but that most researchers in the field would probably agree that it is a system composed of many interacting parts, such that the collective behavior of those parts together is more than the sum of their individual behaviors.'' An important characteristic of complex systems is what are called \textit{emergent behaviors}, the unexpected macroscopic behaviors that often result from simple microscopic interactions between constituent elements \cite{Mitchell2009, Newman2011}. These characteristics---significant interactions and emergent behaviors---collectively suffice to begin defining engineered versions of such systems.\footnote{The term complexity also appears throughout the complex systems literature, along with the field of complexity science. For the purposes of this article, \textit{complexity} refers to the characteristics of a complex system and are thus roughly synonymous with complex systems, though entire texts exist to define the term and field more completely \cite{Mitchell2009}.}

\subsubsection{Complex Engineered Systems}
\label{subsubsec:background_definitions_complexengineeredsystems}

As with complex systems, consensus does not exist about the definition of a \textit{complex engineered system} \cite{Simpson2011}. The simplest definition is ``networks of interconnected components'' \cite{Sosa2011}, though the focus therein limits itself to networks of physical artifacts. Braha et al. \cite{Braha2006} define CES as ``engineered systems [that] are composed of many heterogeneous subsystems and are characterized by observable complex behaviors that emerge as a result of nonlinear spatio-temporal interactions among the subsystems at several levels of organization and abstraction.'' Put another way, they are complex systems composed of designed artifacts with interactivity and potential for emergence, thereby making CES a subset of complex systems.

\subsubsection{Engineering Systems}
\label{subsubsec:background_definitions_engineeringsystems}

So far, the definitions of CES do not consider the significant role that people play is the creation, operation, and evolution of these systems. The term \textit{engineering system} largely derives from the Massachusetts Institute of Technology's former Engineering Systems Division, now part of the Institute for Data, Systems, \& Society \cite{DeWeck2016}, and takes a more social perspective on engineered systems. Their definition, ``a class of systems characterized by a high degree of technical complexity, social intricacy, and elaborate processes, aimed at fulfilling important functions in society,'' \cite{DeWeck2011} underscores the social purposes and objectives while acknowledging the technical, procedural, and social interaction. Much like how information scholars embrace social subsystems of systems via the term socio-technical system \cite{Leonardi2012}, the term engineering system ensures that human interaction with technology and other people remains integral, and consequently unique among the definitions.

\subsubsection{Systems of Systems}
\label{subsubsec:background_definitions_systemsofsystems}

Moving closer to practitioners of engineering, the terminology changes yet again. The DoD provides a formal definition of a \textit{system of systems}, ``a set or arrangement of systems that results when independent useful systems are integrated into a larger system that delivers unique capabilities'' \cite{ODUSD2008}. International standards bodies provide similar definitions \cite{ISO2015}. Papers on systems of systems often refer to such standards \cite{Gorod2008, Ferreira2010} reflecting the industry-orientation of the term and practice-based attempts to address complexity \cite{Norman2006, Clark2009, Baldwin2015, Dahmann2016}, though academics utilize similar definitions \cite{Keating2003}. Valid critiques exist on both sides: Bar-Yam \cite{Bar-Yam2006} notes that traditional systems engineering practices struggle to manage the many interactions and interdependencies between systems, though Alderson \& Doyle \cite{Alderson2010} similarly critique complex systems for not addressing issues of practice.

\subsubsection{Large-Scale Complex Engineered Systems}
\label{subsubsec:background_definitions_laces}

This practice orientation gives way to another definition seeking to bridge theory and practice. \textit{Large-scale complex engineered systems} (abbreviated LSCES or LaCES) are ``engineering projects with significant cost and risk, extensive design cycles, protracted operational timelines, a significant degree of complexity, and dispersed supporting organizations'' \cite{Meluso2018}. The term grew out of a confluence of experts at the NSF, the National Aeronautics and Space Administration (NASA), academia, and industry to address the challenges described in Sec. \ref{sec:introduction} \cite{Deshmukh2010, Lewis2012, Bloebaum2012, Collopy2012, Bloebaum2012a}. Proponents advocated for the development of systems engineering theory \cite{Deshmukh2010, Collopy2012}, of engineering design \cite{Lewis2012}, optimization techniques \cite{Bloebaum2012, Collopy2012, McGowan2013}, and socio-technical understandings of systems \cite{Bloebaum2012, Bloebaum2012a, McGowan2013}. Despite efforts to incorporate the various interests of the academy and industry, the technical and the social, the term remains limited to the mechanical design community.

\subsubsection{A Working Definition of Complex Engineered Systems}
\label{subsubsec:background_definitions_workingdefinition}

Given these definitions and the various concepts they encompass, this paper uses the term \textit{complex engineered system} (still abbreviated CES) defined as large sets of highly-interacting engineered artifacts with a defined purpose. This definition includes the interactivity and potential for emergence of complex systems. While the definition does not explicate human influence, it presumes that people design, operate, and evolve artifacts. Similarly, while environments, contexts, and non-designed objects also interact with the system, they remain peripheral until integrated or modified by people, at which point they too become artifacts. Through this definition it becomes possible to advance theory of how CES develop.

\subsection{Constituents of CES Development Processes}
\label{subsec:background_constituents}


This section is not a description of specific development processes, such as the Systems V \cite{Keating2003, Clark2009} or the Toyota Production System \cite{Adler1999, Spear1999}, but of the abstracted elements that compose a CESDP. These elements include:
\begin{enumerate}[label={(\alph*)}]
    \item the \textit{\textbf{organization}} developing the CES,
    \item the \textit{\textbf{context}} in which the organization develops the CES,
    \item the \textit{\textbf{process}} through which the organization develops the CES, and
    \item the \textit{\textbf{complex engineered system}} itself.
\end{enumerate}
To fully understand how CES develop, researchers and practitioners alike must understand all four of these elements. The following sections address these elements in turn.

\subsubsection{Organizational Theory}
\label{subsubsec:background_constituents_orgtheory}

Management decisions affect organizational performance, and consequently, system performance. This holds true with subsystem complexity \cite{Gokpinar2010}, the balance between project and functional management in a matrix organization \cite{Katz1985}, employee incentives \cite{Natter2001}, managerial responses to failure \cite{Eggers2019}, and even employee perceptions of procedural justice in top management decisions \cite{Li2007}. The complexity of these systems necessitates organizational involvement, so leading scholars argue organizational theory is necessary for creating systems engineering theory as well \cite{Triantis2014} because engineering is a fundamentally social activity \cite{Collopy2012a}.

One of the core organizational theory texts is Scott \& Davis' \textit{Organizations \& Organizing} \cite{Scott2007}. Historically, three views of organizations emerged: as \textit{rational systems} that presume formal relational structures between people who share a common goal; as \textit{natural systems} wherein people have multiple (shared and individual) goals and the relational structure serves as a resource; and as \textit{open systems} through which resources pass and interact with shifting coalitions of people.\footnote{In organizational theory contexts, ``organizations are \textit{systems} of elements, each of which affects and is affected by the others'' \cite{Scott2007} much as in technical systems, though here the term includes people, information, and artifacts.} Each theoretical perspective bears similarity with CES concepts and may serve as foundations upon which to build theory.

\subsubsection{Context}
\label{subsubsec:background_constituents_context}

Defining a system's boundaries is one of the most important tasks in system definition, and no less important are understanding inputs, outputs, and external forces or ``externalities'' \cite{DeWeck2011}. The breadth or narrowness of a project's scope can affect performance outcomes \cite{Gerwin2002}. Just as the context of the system affects its performance, so too does the context of the organization such as through information technology selection \cite{Melville2004}, the influence of marketing on project decisions \cite{Atuahene-Gima2000}, whether the companies involved in development are vertically or horizontally integrated \cite{Kapoor2011}, and organizational culture \cite{Saffold1988}. Theoretical work in this space is primarily limited to the organization and management sciences to date, though it certainly merits further inquiry given its importance.

\subsubsection{Development Processes}
\label{subsubsec:background_constituents_devprocesses}

Practitioners have long suggested that an organization's choice of development process matters, from Kelly Johnson and the Lockheed ``Skunk Works''\cite{Johnson1985} to present-day Silicon Valley start-up culture. Unsurprisingly, system performance depends on the chosen development process \cite{Gerwin2002}. Why is this the case? How do different processes achieve better performance than one another? What development process characteristics make it more or less likely to achieve an organization's strategic objectives for ``simple'' systems let alone more complex ones?

Currently, insufficient theory exists to answer these and similar questions \cite{Collopy2015}. Some knowledge exists in the management sciences about ``new product development'' \cite{Zirger1990, Schilling1998, Schilling1998a, Mulotte2013} and engineering design researchers are conducting experiments on micro-level processes that produce better designs \cite{Austin-Breneman2011, Lewis2012, McComb2017, Eckert2019, Carbon2019}. Still, much remains unknown.

\subsubsection{Theory on Complex Engineered Systems}
\label{subsubsec:background_constituents_CEStheory}

The majority of existing theory on CES grounds itself in network theory. Network scholars observed that the internet, the world wide web, power grids, and other CES could be represented as networks \cite{Irnan1994, Barabasi1999a, Carlson1999, Newman2003a, Clauset2009}. Engineering scholars have characterized the networks of several CES as well including spacecraft, vehicles, and software. Theories have begun to form around structure \cite{Braha2007, Sosa2011}, modularity \cite{Sosa2003, Sosa2007, Sinha2018}, quality \cite{Gokpinar2010, Sosa2011}, innovation \cite{Lungeanu2015}, and the limitations of systems engineering for CES \cite{Bar-Yam2006}. A small contingent of work also studies how uncertainty affects system performance via game theory \cite{Austin-Breneman2015, Bhatia2016, Meluso2018}.

\subsection{Methods for Scientifically Studying CESDPs}
\label{subsec:background_methods}


Building systems engineering theory requires formal methodologies with which to create those theories. Given the elements of CESDPs described in Sec. \ref{subsec:background_constituents}, researchers can employ existing scientific methods used in organizational scholarship to study organizations, contexts, processes, and CES. As with much of social science research, organizational research methods tend to fall under the categories of qualitative and quantitative.

Researchers typically utilize qualitative methods---including interviews \cite{Rowley2012}, ethnography \cite{Ybema2009}, case studies \cite{Gibbert2010}, and grounded theory \cite{Corbin2008}---to answer questions of ``what happens,'' ``how it happens,'' and ``why it happens'' in organizations. Prominent examples with engineering organizations include assessments of the Three Mile Island \cite{Perrow1981} and \textit{Challenger} \cite{Vaughan1997} disasters, though studies also utilize such methods to broaden and deepen understandings of non-catastrophic activities in engineering organizations \cite{McGowan2013, McGowan2014, Austin-Breneman2015, Szajnfarber2017, Meluso2018, Meluso2019}.

From understandings of the ``whats, hows, and whys,'' researchers draw on quantitative methods---surveys \cite{Schuman2002}, content analysis \cite{Duriau2007}, network analyses \cite{Ahuja1999, Newman2018}, simulations \cite{Harrison2007, Morgan2015}, statistical approaches, etc.---to assess the extent, relationships, and relative impacts of phenomena on macro-level outcomes of interest. Note, though, that quantitative data are similarly social in organizational contexts and consequently require care \cite{Schuman2002, DIgnazio2020}.

Unfortunately, gathering data on enough CESDPs to reach statistically significant conclusions is not feasible today due to limits on the volume of publicly-accessible data that describe substantial portions of development processes \cite{Meluso2019}. In such cases, agent-based models (ABMs) and multi-agent systems (MAS) are promising tools for assessing the macro-level effects of micro-level behaviors \cite{Bruch2013, Macal2016}.\footnote{The difference between ABMs and MAS is subtle. ABMs typically emphasize \textit{descriptive} understandings of the interactions between agents whereas MAS take \textit{normative} approaches to coordinate agents toward the best collective outcome \cite{Macal2016}. Both have their place in CES research depending on the scientific objectives.} Through either of these techniques, researchers can simulate communication \cite{Meluso2019}, game theoretic interactions \cite{Axelrod1997, Bhatia2016}, and numerous network dynamics \cite{Newman2018}. Standard Design of Experiments (DOE) techniques apply for understanding both the characteristics under study and the parameters of the model \cite{Sacks1989, Sanchez2002, Lee2015}.\footnote{See Sec. \ref{subsubsec:background_CESIUM_ABMs} for more on ABMs.}

\subsection{Methods Specific to CESIUM}
\label{subsec:background_influences}

The availability of diverse methods creates opportunities for combining methods into representations of CESDPs, as with CESIUM. The following sections describe the contributions of network theory, agent-based modeling, and design optimization to CESIUM. \footnote{Sec. \ref{subsec:background_influences} was partially adaptated from Sec. II-B of \cite{Meluso2019}.}

\subsubsection{Network Models}
\label{subsubsec:background_CESIUM_networks}

Network theory represents systems of people or artifacts as nodes and edges, as shown in Fig. \ref{fig:network_graph}. \textit{Nodes}, points connected to one another in pairs, may represent people, subsystems, artifacts, etc. in a CES. Connections or interactions between nodes are called \textit{edges} or \textit{ties} and can be represented by an adjacency matrix $A_{ij}$, where $A_{ij}=1$ if an edge exists between nodes $i$ and $j$ and $A_{ij}=0$ otherwise \cite{Newman2018} as in Fig. \ref{fig:adjacency_matrix}. When thought of as interfaces between artifacts, a $A_{ij}$ forms a Design Structure Matrix representing a CES \cite{Sosa2011}; thought of as a social network, edges represent pathways for information diffusion \cite{Jiang2015}. The edges of an adjacency matrix may be either directionless (called \textit{undirected} edges) or \textit{directed} from one node to another, in which case $A_{ij}=1$ only if an edge points from node $j$ to node $i$ \cite{Newman2018}.

\input{figures/figure_network}

The number of other artifacts that each artifact $i$ interfaces with is called the degree $k_i$ of $i$. A normalized histogram of a network's degrees is called a \textit{degree distribution} \cite{Newman2018}. Ample studies have shown that artifacts in many (but not all) complex systems follow a \textit{scale-free} degree distribution, also called power-law or inverse exponential distributions \cite{Barabasi1999, Albert2002, Newman2003, Braha2004, Braha2007, Clauset2009, Sosa2011}. Scale-free distributions take the form $p_k=c_1 k^{-c_2}$ where $p_k$ is the probability of randomly selecting a node with degree $k$, $c_1$ is a constant, and positive constant $c_2$ is the exponent of the power law with typical values of $2\leq c_2\leq3$ \cite{Newman2018}. The resulting function appears as a negatively-sloping line in a log-log plot as in Fig. \ref{fig:degree_distribution}. Braha \& Bar-Yam \cite{Braha2004, Braha2007} and Sosa et al. \cite{Sosa2011} suggest that complex system degree distributions generally follow a power-law with a cut-off at some large degree reminiscent of a bow-tie structure, making systems resilient to most small perturbations but vulnerable to certain rare perturbations \cite{Carlson1999, Kitano2004}. Consequently, scale-free and bow-tie distributions may each provide unique insights.

A more-recent subject in network theory with significant potential is that of generative network models that algorithmically construct networks out of basic rules \cite{Newman2018}. One of the most common network generation algorithms is called preferential attachment which builds a network by connecting new nodes to existing nodes with an attachment probability proportional to the degree of the existing node \cite{Newman2018}. Several such algorithms exist including those of Price \cite{Price1976}, Barabási \& Albert \cite{Barabasi1999}, Holme \& Kim \cite{Holme2002}, and Carlson \& Doyle \cite{Carlson1999}, all of which generate networks with scale-free degree distributions \cite{Newman2018, Holme2002}. Recent advances in peer-to-peer network studies allow generative algorithms to establish hard or soft cut-offs in the distribution \cite{Guclu2007, Kumari2011}. The Holme-Kim preferential attachment algorithm includes a parameter for tuning node clustering \cite{Holme2002}, making it useful for simulating the meso-scale practice of subsystem formation. On the other hand, Carlson \& Doyle's highly optimized tolerance (HOT) generates bow-tie structures, yielding macro-level realism. An approach combining the Holme-Kim and HOT approaches would likely provide the most realistic CES structures, albeit similarly presumptive of solution structures.

\subsubsection{Agent-Based Modeling}
\label{subsubsec:background_CESIUM_ABMs}

Agent-based modeling is a widely-used, effective, and tested method for simulating CES \cite{Grimm2006, Grimm2010, Bonabeau2002, Macal2010}. An ABM creates a system of autonomous decision-making entities called agents which individually assess their situations and make decisions based on a set of rules \cite{Bonabeau2002}. Agents affect their surroundings through their actions and in doing so, self-organization, patterns, structures, and behaviors emerge from the ``ground up'' that were not explicitly programmed into the models but nevertheless arise through agent-interactions \cite{Macal2010}. This ``ground up'' agent-centered approach differentiates ABMs from other system modeling methods such as discrete event simulation and system dynamic models which take top-down approaches \cite{Macal2016}. The ability of ABMs to demonstrate emergence also makes it ideal for understanding highly-interacting systems.

Recent applications of ABMs include systems design \cite{Panchal2009, Darabi2017, Soyez2017, Meluso2018, Meluso2019} and organization studies \cite{Carley1996a, Carley1997, Anjos2013, Jamshidnezhad2015, Meluso2019}. INCOSE promotes ABMs as one of the primary methods through which ``to inform trade-off decisions'' regarding ``complexity in system design and development'' \cite{Salvucci-favier2016}. Because CES are often composed of many smaller engineered systems that are designed, developed, and operated by organizations of dispersed, loosely connected people \cite{Bloebaum2012}, ABMs facilitate simulation of aggregated artifact development in ways that top-down models cannot \cite{Macal2016}.

\subsubsection{Design Optimization}
\label{subsubsec:background_CESIUM_MDO}

Engineers in various disciplines use \textit{design optimization} to maximize the performance of a system, a process of selecting the relative ``best'' alternative from among a set of possible designs called the \textit{design space} \cite{Papalambros2017}. They do this through \textit{objective functions} (or \textit{utility functions} in agentic contexts), sets of evaluation criteria typically constructed as functions describing the relationships between independent or \textit{decision variables} \cite{Papalambros2017}. Optimization algorithms then explore the design space to find a global or local minimum (or maximum depending on problem construction) as efficiently as possible to identify a solution \cite{Martins2013}.

While the methods of constructing system objectives are beyond the scope of this paper, one method for searching design spaces remains relevant. Validated studies have shown that engineers \cite{McComb2015, McComb2016} and organizations \cite{Carley1996a} sample their design spaces comparably to \textit{simulated annealing} which can therefore serve as a modeling representation of human decision-making.

Given this background, the next section combines these concepts to form CESIUM through the framework established in Sec. \ref{subsec:background_constituents}.

\section{Methodology}
\label{sec:description}

The CESIUM framework simulates a theoretical CESDP by modeling its constituent elements (Sec. \ref{subsec:background_constituents}). This section describes how the \textit{base instance} of CESIUM represents each element: the system architecture and system boundary (Sec. \ref{subsec:methodology_system_architecture}), the constituent artifacts (Sec. \ref{subsec:methodology_artifacts}), the organization of interacting agents (Sec. \ref{subsec:methodology_organization}), agents' design processes (Sec. \ref{subsec:methodology_agent_process}), and the system development process (Sec. \ref{subsec:methodology_system_development}). Thereafter, an example execution of the base instance follows for a vizualizable system (Sec. \ref{subsec:methodology_example}) along with a brief introduction to the framework's flexibility in simulating variations on the elements of CESDPs (Sec. \ref{subsec:methodology_flexibility}).\footnote{The implementation of CESIUM described herein was developed using Python 3 with the NumPy 1.18.1, SciPy 1.4.1, and NetworkX 2.4 packages \cite{Harris2020, Virtanen2020, Hagberg2008}. The full code for this article is available at \url{https://github.com/meluso/cesium-framework}, and a clean version of code for the base instance of CESIUM is available at \url{https://github.com/meluso/cesium-base}.}\footnote{Secs. \ref{subsec:methodology_system_architecture}--\ref{subsec:methodology_system_development} were partially adapted from Sec. III of \cite{Meluso2019}.}


\subsection{System Architecture}
\label{subsec:methodology_system_architecture}

First, assume that a CES is composed of $n$ interacting artifacts where each artifact $i \in \{1,\ldots,n\}$. The $n$ artifacts interact with one another in a technical network generated from a scale-free degree distribution via a Holme-Kim preferential attachment algorithm by adding $h$ edges to each new node $i$. The probability that the first edge connecting node $i$ to the network will attach to a specific node $j$ is proportional to that node's degree $k_j$; the probability that subsequently-added edges will be placed to form a triangle by connecting $i$ to a node $l$ that is already connected to $j$ is $p_t$ \cite{Holme2002}. The result is a network of $n$ interacting artifacts described by adjacency matrix $A_{ij}$, in this case with no formal hierarchy and clustering specified by $p_t$. Generally, artifacts could also interact with nodes outside the system boundary. The base instance assumes that the system boundary contains all factors with sufficient impact on the CES leaving no exogenous variables, and hence no interaction between the CES and its context.

\subsection{Artifact Construction}
\label{subsec:methodology_artifacts}

\input{figures/figure_interaction}
\input{figures/figure_objectives}

In a real-world setting, the design of each artifact $i$ in the system would depend on numerous contextual and specific factors, say $\Vec{v}_i=[v_{i1},v_{i2},\ldots]$. Because these factors cannot be known \textit{a priori} for innumerable real systems, the model representatively parameterizes these variables such that the design of each artifact is defined by a single decision variable $x_i(\Vec{v}_i)$. Therefore, each $x_i$ parameterizes a complex set of inputs, allowing the performance of each artifact to be represented as an objective function $y_i=f_i(x_i,\Vec{x}_j)$, where $j\in\{1,\ldots,k_i\}$ represents the set of artifacts interfacing with artifact $i$, and $\Vec{x}_j$ is a vector of the parameterized decision variables of the $k_i$ artifacts as exemplified in Fig. \ref{fig:interaction}. These variables can be expressed using a combined notation $\Vec{x}_i=[x_i,\Vec{x}_j]$. 

Objective functions can take countless forms in the framework, but because $\Vec{x}_i$ can be parameterized to any mapping, the simplest objective topologies can be identified through a Taylor expansion of $f_i(\Vec{x}_i)$. At some point $\Vec{x}_{i0}$, with $\Vec{\partial x}_i=\Vec{x}_i-\Vec{x}_{i0}$ and Hermitian matrix $H_i(\Vec{x}_i)$, the Taylor expansion of $f_i(\Vec{x}_i)$ is
\begin{equation}
    \begin{split}
        f_i(\Vec{x}_i) \approx f_i(\Vec{x}_{i0}) & + \nabla f_i(\Vec{x}_{i0}) \Vec{\partial x}_i \\
        & + \frac{1}{2} \Vec{\partial x}_i^T H_i(\Vec{x}_{i0}) \Vec{\partial x}_i + \cdots
    \end{split}
\end{equation}

If $\Vec{x}_i$ is parameterized such that the optima all occur at $\Vec{x}_i^*=\Vec{0}$, then $f_i(\Vec{x}_i\neq\Vec{0})>f_i(\Vec{x}_i^*=\Vec{0})$ for every $i$. The simplest such relationship occurs under the condition of linearity such that $|\nabla f_i(\Vec{x}_{i0})\Vec{\partial x}_i|=\sum_{m=1}^{k_i+1} |x_m|$ with all higher order terms equal to 0. The next simplest then is a quadratic formed by taking only the second term of the Taylor expansion such that $|\frac{1}{2} \Vec{\partial x}_i^T H_i(\Vec{x}_{i0}) \Vec{\partial x}_i| = \sum_{m=1}^{k_i+1} x_m^2$. More complex relationships, with multiple minima and asymmetrical objective functions, will of course prove more realistic though may prove more challenging to formulate and analyze. However, those complexities can similarly be overcome through the initial parameterizations of each $x_i$.

Hence, to capture a variety of CES, the base instance considers four objective functions that correspond to the first- and second-order Taylor expansion terms, a symmetrical multiple-minima function, and an asymmetrical multiple-minima function as follows:
\smallskip
\begin{enumerate}[label={(\alph*)}]
\item \input{figures/eq_absolute-sum}
\item \input{figures/eq_sphere}
\item \input{figures/eq_ackley}
\item \input{figures/eq_levy}
\end{enumerate}
\bigskip
Fig. \ref{fig:objective_functions} shows these functions in two dimensions.

One of the advantages of these particular objective functions is that they are $n$-dimensional, meaning they scale to incorporate the $k_i$ decision variables for each neighbor $j$ of $i$ for any positive integer value of $k_i$. This leaves $n$ coupled objective functions $\{f_1,\ldots,f_n\}$ that comprise the CES being designed.

\subsection{The Organization}
\label{subsec:methodology_organization}

CESIUM assumes that the members of an organization (engineers or otherwise) can be represented by agents in an agent-based model. In the base instance, one agent represents one engineer. While, the pairings of agents to artifacts can and do take many forms in real-life, an organization's structure approximately reflects the structure of the technical artifacts that those organizations create. This phenomena, called the mirroring hypothesis or Conway's Law \cite{Cabigiosu2012, MacCormack2012, Colfer2016}, means the simplest mapping is one in which the network of agents in the organization and the network of artifacts in the CES are synonymous as assumed herein. Each agent is then responsible for one artifact in the system. As a result, an organization exists wherein engineers pass information via the technical network.\footnote{Sec. \ref{subsec:future_framework} discusses more complex mappings, including variation in the numbers of agents artifacts, as part of the generalization of CESIUM.}

\subsection{Agent Design Process}
\label{subsec:methodology_agent_process}

Next, the model incorporates a design process for the artifacts. Given the mirroring hypothesis, each agent uses the technical objective function of its artifact as its utility function, so the objective functions will be spoken of as belonging to the agents. Each agent seeks to optimize (that is, minimize) its objective function over a number of turns to reach the best performance.\footnote{While stepping forward in time through the use of turns is a common practice in ABMs, the system-level motivation and analog for turns are described in Sec. \ref{subsec:methodology_system_development}.}

Again, validated studies have shown that engineers sample their design spaces similar to optimization using simulated annealing \cite{McComb2015, McComb2016, Carley1996a}. During each turn of the model, each agent engineer receives a set of constant input values $\Vec{x}_j$ from its $k_i$ interacting agents to utilize when adjusting $x_i$ to optimize their objective functions. Agents explore the design space using a simulated annealing algorithm in search of a local optimum $y_i^*=f_i(x_i^*,\Vec{x}_j)$ with a random initial position in the domain of $x_i$, $\omega$ iterations per optimization, initial temperature of $\tau$, and cooling rate of $\rho$. No cognitive factors affect agent decision-making. While other design space search algorithms remain possible, the base instance of CESIUM uses simulated annealing for agent optimization due to its established validation.

\subsection{System Development Process}
\label{subsec:methodology_system_development}

With a technique established through which members of an organization develop each artifact, it becomes feasible to simulate the development of the complete system. Returning to the beginning of the CESDP: the ABM first initializes a new system following the method outlined in Sec. \ref{subsec:methodology_system_architecture}. Then, the model steps through a series of \textit{turns} where one turn in the ABM represents one design cycle in a CESDP. From one perspective, turns represent a common development process technique known as the Shewhart \& Deming Cycle \cite{Anderson1994} wherein members of an organization iteratively improve and share the design for their subset of a system; though, turns also serve as a discretized representation of design refinement more generally.

Each turn, agents exchange information. To facilitate the exchange, CESIUM stores the latest reported designs of all agents in a system vector $\Vec{S}$ as a central repository. At the beginning of each design cycle, each agent receives $\Vec{S}$ as a constant input before proceeding to optimize their variable $x_i$ using only the values from their networked neighbors $\Vec{x}_j$. Then, each agent passes their updated value of $x_i$ back to the system vector for storage in $\Vec{S}$ and a new design cycle begins with the updated values as constants.

The model performs these design cycles, iterating through all of the agents in each cycle, until either the system design converges or the model performs $d$ design cycles. System convergence is calculated from a metric for System Performance $F$. While many formulations of performance are possible, the base instance defines $F$ as a sum of the reported objective evaluations of all of the agents during the current design cycle:
\begin{equation}
    F(t,\Vec{f})=\sum_{i=1}^n f_i(t,\Vec{x}_i(t))
\end{equation}

Although the $n$ objective functions $f_i$ have different magnitudes depending on the degree $k_i$ of each artifact $i$, the System Performance is assumed to have a greater dependence on components which are more highly connected so simply adding their contributions exemplifies this behavior. By this definition, system convergence occurs when the System Performance consistently changes less than a specified convergence threshold $\varepsilon$ over each of three turns, that is, meeting the following condition:
\begin{equation}
    \frac{1}{3}\sum_{t'=1}^{3} |F(t,\Vec{f})-F(t-t',\Vec{f})| < \varepsilon
\end{equation}

Every execution of the model completes a minimum of three design cycles. Upon convergence, the Number of Design Cycles $N$ measures how long it took the organization to converge to a design solution, and therefore to complete the CESDP. This completes the base instance of CESIUM.

\subsection{Example Execution}
\label{subsec:methodology_example}

\input{figures/figure_example}

To illustrate this complete development process simulation, this section provides an example execution for a small system which demonstrates how CESIUM simulates individual and an organizational exploration of their respective design spaces with varying effects on system outcomes. Because larger numbers of agents become difficult to visualize, consider a small example system of the base instance with only two agents as in Fig. \ref{fig:example_execution}. Each of the $i\in\{1,2\}$ agents controls one decision variable, $x_1$ and $x_2$, respectively. Assuming that the artifacts are neighbors in the system network, consider the case where agents' objective functions $f_1(x_1,x_2)$ and $f_2(x_2,x_1)$ are Levy functions, as in Fig. \ref{fig:fn_levy}, with System Performance $F(t,f_1,f_2)$.

Unlike the other objectives presented herein, the Levy function is not symmetric meaning that some $f_i(x_i,x_j)$ is not necessarily equal to $f_i(x_j,x_i)$. Given the ordering of the variables for each agent's objective function, Agent 1 has limited control over its outcomes because changes in $x_1$ tend to yield smaller changes in $f_1$ than changes in $x_2$ due to the shape of the Levy function, and vice versa for Agent 2. This configuration then represents a system in which the performance $f_i$ of each artifact $i$ is disproportionately affected by design decisions made for the $k_j$ artifacts that artifact $i$ interacts with through the system network making convergence to the global optimum more challenging.

Figs. \ref{fig:example_f1} \& \ref{fig:example_f2} show the progress of each agent toward its individual objective as a function of the two decision variables, and Fig. \ref{fig:example_system} shows the system convergence and progress toward the global optimum. Observe how changes in $x_2$ tended to improve $f_1$ even as they produced limited improvement in $f_2$; however, those changes in $x_2$ eventually led $x_1$ to identify improved values of $f_1$ and consequently $x_2$ to do the same for $f_2$. The iterative process of simulated annealing---a combination of necessarily accepting values of $x_i$ that improve $f_i$ and probabilistically accepting values of $x_i$ that do not improve $f_i$---gradually improved System Performance as in Fig. \ref{fig:example_system}. Hence, this example execution demonstrates how CESIUM simulates both individual and organization exploration of a design space toward collective system outcomes.

\subsection{Framework Flexibility}
\label{subsec:methodology_flexibility}

Throughout Secs. \ref{subsec:methodology_system_architecture}--\ref{subsec:methodology_system_development}, the base instance makes a number of simplifying assumptions. However, the CESIUM framework \textit{requires} very few of these. Indeed, only the elements described in Sec. \ref{subsec:background_constituents} \& forthcoming Sec. \ref{subsec:future_framework} need accounting for, thereby creating the flexibility for researchers to alter any other qualities described herein so as to enable exploration of other characteristics, from system architecture topologies to representations of established development processes.

For example, Meluso et al. \cite{Meluso2019} recently utilized the framework to simulate miscommunication in a CESDP. Building on prior findings that practitioners define the ``estimates'' they communicate with one another either as representations of their \textit{current} design status or \textit{future} design outcome \cite{Meluso2020}, the authors introduced a variable $p_e$ specifying the probability that an agent would share future estimates (or conversely $1-p_e$ that they would share current estimates) finding that miscommunication can affect system performance. Such investigations readily follow from Sec. \ref{sec:future}, which generalizes the modeling framework and recommends further avenues of inquiry.

\section{Model Analysis}
\label{sec:analysis}

\input{figures/table_parameters}
\input{figures/table_reg_perf}
\input{figures/table_reg_cycl}
\input{figures/figure_feat_importances}

Model design decisions affect outcomes with any model. But as this section will demonstrate, the strength of CESIUM as a modeling framework lies in its ability to expose patterns that emerge across variation in design decisions. Abstracting away from particular CESDPs to their constituent elements (Sec. \ref{subsec:background_constituents}) creates the opportunity to explore the fundamental characteristics of those elements, such as the number of artifacts in a system, the resources provided to engineers, or the qualities of the phases of a development process.

To that end, this section fills three purposes. First, Sec. \ref{subsec:analysis_sensitivity} characterizes the model through sensitivity analysis. In so doing, the analysis reveals a tremendous range of possible outcomes contingent on variables representative of practical system and design process qualities. Then, Sec. \ref{subsec:analysis_validation} validates the framework. Sec. \ref{subsec:analysis_theory} closes by demonstrating the modeling framework's power to identify testable hypotheses and postulate theories describing how real phenomena in practice affect system outcomes.

\subsection{Sensitivity Analysis}
\label{subsec:analysis_sensitivity}

The authors performed a sensitivity analysis \cite{TenBroeke2016} to characterize the performance of CESIUM with different representations of design processes by sweeping independent variables, examining their main and interaction effects via regression, and determining overall feature importances with a random forest.


Table \ref{tab:parameters} contains the independent variables, constants, and dependent variables; their definitions; and the values varied as part of a fractional factorial experiment. The subset of independent variables were selected because
\begin{enumerate*}[label={(\alph*)}]
    \item they easily map to real-world concepts (e.g. system size, development time), and
    \item they were assessed most likely to produce variation in the dependent variables.
\end{enumerate*}
Sec. \ref{subsec:methodology_flexibility} also mentioned the Future Estimate Probability  $p_e$---the probability that agents will exchange estimates of their future design projections instead of their current design values \cite{Meluso2019, Meluso2020}. Variable  $p_e$ was included to demonstrate how deviating from assumptions can reveal more and less optimal strategies for CESDPs contingent upon other variables (see Sec. \ref{subsec:analysis_theory}). Combined, the independent variable sets consisted of 13,552 unique combinations, each of which ran 100 times for a total of 1,355,200 executions of CESIUM.

The authors first quantified the main and interaction effects by utilizing multivariate ordinary least squares (OLS) regressions with robust standard errors \cite{TenBroeke2016,James2013}, as described by Models II.1--7 in Table \ref{tab:regression_performance} for system performance $F$, and by Models III.1--6 in Table \ref{tab:regression_cycles} for the number of design cycles $N$. To start, Models II.1 \& III.1 found main effects for the five independent variables through OLS regression of the full dataset (Num. Obs. $=$ 1,355,200) with objective function dummy variables capturing the objective function fixed effects and the Absolute Sum function represented in the constant. Model II.1 accounted for 61.8\% of the System Performance variance and Model III.1 for 72.4\% of the variance in the Number of Design Cycles. These two models suggested that CESDP outcomes significantly depend on design space objective functions, system size (Num. Nodes.), development process stopping conditions (Conv. Thresh.), and even the type of information exchanged within development processes (Fut. Est. Prob.). Still, the large residual standard errors indicated significant variation remained unaccounted for which could reside in interaction or more complex relationships.

Models II.2 (R-squared: 96.9\%) \& III.2 (R-squared: 81.3\%) included interaction effects, moderately improving the residuals over Models II.1 \& III.1 while attributing significant variation to most of the interaction terms. The significant differences between the objective function interaction coefficients supported that an organization's ability to develop an optimal CES significantly depends on the objective functions corresponding to the design spaces of its constituent artifacts. Models II.3--6 and III.3--6 provided further support for this hypothesis by isolating the interaction terms and residual standard errors wthin the objective function data subsets (Num. Obs. $=$ 338,800 each). The Levy function absorbed the largest residuals, leaving more-proportionate residuals for the other functions while measuring a variety of interaction effect amplitudes within the functions. The Sphere function System Performance (Model II.4) remained an outlier among the regressions, though, only explaining 12.5\% of the variance before log transformations of the System Performance, Number of Nodes, and Convergence Threshold (Model II.7, R-squared: 66.3\%).

According to the models, nearly all of the system outcomes depend strongly on the Number of Nodes, the Convergence Threshold, or both, though rarely through interaction effects. The Number of Nodes tended to increase the Number of Design Cycles, while the Convergence Threshold consistently revealed an inverse relationship between System Performance and the Number of Design Cycles, regardless of the objective function selected. In contrast, the main and interaction effects of the Triangle and Future Estimate Probabilities proved scattered with only select cases generating significant effects. For example, Models II.3 (Absolute Sum), II.4 (Sphere), \& II.5 (Ackley) identified statistically significant effects between the probability of triangle formation in the system network and the probability of sharing future estimates (Tri. Prob. $\times$ Fut. Est. Prob.) with magnitudes comparable to their respective constants, an effect that was limited in the Ackley and absent from the Levy function.


To further support the previous regressions, and capture any non-linear or higher-order effects of each variable, a feature importance analysis using a random forest regressor was applied, shown in Fig.~\ref{fig:feat_importances}. Random forests fit an ensemble of decisions trees to predict the outcome variable, and feature importances describe which variables best reduce errors in outcome predictions when trees split the data on that variable~\cite{breiman2001a}. Random forests were fit using the default parameters of sci-kit learn v0.24.1 with a maximum depth of four. The results proved largely consistent with the regressions with a few noteworthy exceptions. Again, the Number of Nodes and Convergence Thresholds proved most important. However, the Triangle and Future Estimate Probabilities only substantially affected the Sphere function performance, suggesting their effects may be limited to particular topologies. This somewhat contradicts the regression results which additionally identified an interaction effect for the Absolute Sum and Ackley functions, suggesting that the regressions for those functions may have missed nonlinear and/or higher-order effects, or excluded uncaptured variation in either the network structure or simulated annealing searches. That said, the contributions of the Triangle and Future Estimate Probabilities remained significant for the Sphere function.

\subsection{Validation}
\label{subsec:analysis_validation}

Because CESIUM models organizational processes, this work draws from organization science for its definitions of micro- and macrovalidation. Microvalidation ensures that the behaviors and mechanisms coded into the model sufficiently represent, or are ``grounded'' in, their real-world analogs. Macrovalidation is divided into face validation---whether results are superficially distinguishable from real-world conceptual phenomena---and empirical validation---whether results are statistically calibrated to match real-world data \cite{Carley1996, Wilensky2015}.

To satisfy the conditions of microvalidation, Sec. \ref{sec:description} grounds each construction decision of CESIUM in either established research or simplified abstractions of real-world practices. To satisfy face and empirical macrovalidation, simulation results would need to correspond to outcomes one might encounter in practice. While face validation occasionally proves possible, Sec. \ref{subsec:background_methods} cites how insufficient data exists about CESDPs to perform empirical validation. Indeed, CESIUM is designed to address the absence of sufficient data by reconstructing the underlying mechanisms governing development processes and generating data instead.

Consequently, it becomes beneficial to view modeling frameworks like CESIUM from a different perspective. Carley refers to computational models as ``hypothesis generating machine[s]'' \cite{Carley1996}, in CESIUM's case generating data with which to form hypotheses about causal variables in CESDPs. With sufficient evidence from multiple sources of knowledge, researchers can validate those hypotheses and they become established systems engineering theories. Thereafter, systems engineers can devise methods for improving the outcomes resulting from those relationships.

Contributing to that process, the following section will demonstrate both the creation of hypotheses and formation of theory by validating the results of the sensitivity analysis.

\subsection{Theory Building Demonstration}
\label{subsec:analysis_theory}

Consider the results of the sensitivity analysis for each independent variable:
\begin{enumerate}[label=\textbf{{Result \arabic*:}},ref=\arabic*,align=left]
    \item The Number of Design Cycles tended to increase proportional to the \textbf{\textit{Number of Nodes}}, while System Performance varied depending on the objective function. 
        \label{result:num_nodes}
    \item Both system outcomes varied depending on the \textbf{\textit{Objective Functions}}, via both main and interaction effects.
        \label{result:obj_fn}
    \item \textbf{\textit{Triangle Probability}} had small or statistically insignificant effects on system outcomes for all but the Sphere function, where System Performance varied as a function of Triangle Probability, Future Estimate Probability, and interaction between the two.
        \label{result:tri_prob}
    \item Increasing the size of \textbf{\textit{Convergence Threshold}} tended to decrease the Number of Design Cycles while degrading the System Performance.
        \label{result:conv_thresh}
    \item Increasing the \textbf{\textit{Future Estimate Probability}} had small or statistically insignificant effects on system outcomes for all but the Sphere function, where System Performance varied through interaction with the Triangle Probability.
        \label{result:fut_est_prob}
\end{enumerate}

Beginning with Result \ref{result:num_nodes}, this result suggests that even if System Performance varies differently for different objective functions, (larger) systems with more artifacts will tend to take longer to converge on system designs. This observation can be posed as a hypothesis for further examination.

\begin{hypothesis}\label{hyp:num_nodes}
    The time it takes an organization to converge on a system design increases as a function of system size.
\end{hypothesis}

Scholars can now examine this hypothesis against existing and novel studies to determine whether it merits conversion into a theory.

Next, note that Results \ref{result:num_nodes}-\ref{result:tri_prob} \& \ref{result:fut_est_prob} involve variation in system outcomes depending on the objective functions of the agents. This, too, could easily form a hypothesis. However, the finding is consistent with extensive research from Game Theory, Decision Science, and forms the underlying premise of Value-Driven Design \cite{Collopy2011}, thereby providing at least face validation for the hypothesis. With abundant evidence from multiple disciplines, the claim is more accurately stated as an established piece of knowledge in the form of a theory. Recall that CESIUM assumes that the utilities of the agents designing system artifacts are synonymous with the objective functions of the artifacts. Then:

\begin{postulate}\label{post:objective_fn}
    The outcomes of a complex engineered system development process depend on the objective functions of the agents who develop system artifacts.
\end{postulate}

Variation in the other regression coefficients coincident with the objective functions leaves question as to the relationship between even the most significant of the remaining variables and system outcomes. Nevertheless, CES outcomes certainly depended on the selected function giving this first postulate in combination with existing evidence.

The results lend themselves to two further hypotheses meriting examination. Result \ref{result:conv_thresh} suggests:

\begin{hypothesis}
    Given a system objective function in negative null form, system performance varies inversely proportionately to system convergence time as a function of maximum uncertainty required in system performance.
\end{hypothesis}

Finally, recall that the Triangle Probability is a proxy for modularity or clustering in the technical network of the system architecture, and that the Future Estimate Probability examines the likelihood that an agent will communicate information about either current design statuses or projected design outcomes gives the final hypothesis. Combining these abstractions with Results \ref{result:tri_prob} \& \ref{result:fut_est_prob} gives:

\begin{hypothesis}
    System performance varies as a function of interactions between modularity in the system architecture and the temporal references of information exchanged within the organization.
\end{hypothesis}

\section{Future Directions}
\label{sec:future}

As a framework, CESIUM's conceptualization of CESDPs presents numerous opportunities for further exploration including system architectures, system development processes, artifact design, patterns and contents of interaction, and objective functions. To facilitate exploration, this section begins by generalizing the framework (Sec. \ref{subsec:future_framework}). Then, it outlines recommended directions for future inquiry based on the framework to facilitate the development of theory (Sec. \ref{subsec:future_recommendations}).

\subsection{Framework Generalization}
\label{subsec:future_framework}

The base instance of CESIUM makes simplifying assumptions about CES and CESDPs including a scale-free degree distribution of artifacts, no exogenous variables, one agent per artifact via the mirroring hypothesis, simulated annealing artifact design, a design cycle development process, and a simple sum of artifact performances as representative of CES performance. The following paragraphs relax those assumptions and, in so doing, opens opportunities for exploration.

Assume that a CES is composed of $n_a$ artifacts interacting in a technical network and system architecture described by adjacency matrix $A_{ij}$. The $n_b$ members of an organization (engineers and otherwise) can be represented by agents in an agent-based model which exchange information with one another in a communication network described by adjacency matrix $B_{qr}$. Each of the $n_b$ agents acts upon or ``works on'' some set of the $n_a$ artifacts according to matrix $C_{iq}$. Summing over $q$ then yields the number of agents that contribute to the design of each artifact $\lambda_i=\sum_q C_{iq}$.

Agents apply design decisions to each artifact they work on, influenced by numerous specific and contextual factors. Describe the decisions made by agent $w\in\{1,\ldots,\lambda_i\}$ about the $i$\textsuperscript{th} artifact with a decision variable vector $\Vec{v}_{iw}=[v_{iw1},v_{iw2},\ldots]$ of contributions made by each agent working on the artifact. The contributions of any variables exogenous to the system can be described by $\Vec{u}_{i}=[u_{i1},u_{i2},\ldots]$. As before, these variables can be representatively parameterized so that each artifact is modeled by a single decision variable $x_i(\Vec{v}_{i1},\ldots,\Vec{v}_{i\lambda_i},\Vec{u}_i)$ (but this time of vectors $\Vec{v}_{iw}$ and $\Vec{u}_i$), thereby mapping a set of complex factors to a more manageable form.

This parameterized specification for the design process of each artifact makes it possible to describe the performance for each as $y_i$ which is a function of the artifact's own design $x_i$, the vector of designs $\Vec{x}_j$ corresponding to the $k_i$ interacting artifacts, and a vector of any exogenous variables $\Vec{z}_i$ that influence performance. The combined notation $\Vec{x}_i=[x_i,\Vec{x}_j,\Vec{z}_i]$ then allows the artifact's performance to be described in terms of an objective function $y_i=f_i(\Vec{x}_i)$. Over time, agents can individually act toward the improvement of their own objective functions, CES objective functions, or both. Hence, given one function for each artifact and another vector of exogenous contributions to the system $\Vec{\xi}$, multi-objective functions can be formed quantifying overall outcomes for the CESDP including system performance at different points in time, $F(t, f_1(\Vec{x}_1),\ldots,f_i(\Vec{x}_i),\ldots,f_n(\Vec{x}_n),\Vec{\xi})$. 

While abstract in this form, these generalizations enable researchers to vary assumptions within the CESIUM framework, from which it becomes possible to explore variable dependencies, pose hypotheses, and form theories about CESDPs. The following section enumerates some such options.

\subsection{Recommended Investigations}
\label{subsec:future_recommendations}

Real technical network distributions vary, so exploring bow-tie structures generated by HOT (Sec. \ref{subsubsec:background_CESIUM_networks}) and complete graphs (in which every node is connected to every other node) could yield novel insights about how artifact interaction networks shape system outcomes. Again, the clustering parameter of the Holme-Kim algorithm leaves it well-suited to study subsystem formation and interaction, closely connecting it to current practices. Of course, artifact networks are rarely static in practice, so incorporating network evolution into CESIUM could shed light on how temporal variation affects both agent-level and system-level outcomes. And because even CES vary dramatically in size, exploring different scales may reveal how outcomes manifest as a result of hierarchical decomposition of systems and system of systems interactions.

Development processes themselves also merit exploration including different permutations of the relationships between artifacts, variable composition, agent contribution, and design space exploration. Artifacts undergo different development processes depending on the industry, design process, designer identity, and designer context \cite{Chou2020}, each of which could be studied through CESIUM. Likewise, many CESDPs employ processes that are more involved than the turn-based Shewhart \& Deming Cycle (Sec. \ref{subsec:methodology_system_development}), including the Systems V, agile development, scrum, etc. The ability to explore such varied development processes may prove one of the greatest opportunities afforded by CESIUM.

The framework also opens possibilities for studying the contents of interactions between contributors. While researchers have begun to investigate communication frequency in other contexts, the contents of communication and interpretation have received limited treatment to date, thereby lending themselves to questions of feedback, coordination between designers \cite{Wen2020}, and different mediums of information exchange.

Different objective constructions could prove extremely informative at the agent, subsystem, system, and system of systems levels. The functions selected for this work began to explore concepts such as linear relationships between variable contributions (Absolute Sum), parabolic relationships (Sphere), local optima (Ackley), and disproportionate influence by collaborators (Levy), though innumerable topologies remain possible at each level. System performance certainly takes forms other than sums of its parts, creating opportunities to study questions of collective innovation, high-reliability organizations, and evolving landscapes, from completely abstract to precisely representative of real-world systems. Hopefully these suggestions will contribute to the beginning of theory building for CES and CESDPs.

\section{Conclusion}
\label{sec:conclusion}

Complex engineered systems play essential roles in society. However, decades of increasing system complexity and development process inadequacies have garnered substantial concern about the capabilities of established systems engineering processes. Leading experts from government, academic, and industry organizations now call for the development of theories to guide solutions to these growing needs. But to date, the notorious difficulty of gathering data from real-world complex engineered system development processes has stymied efforts to form and validate generalizable hypotheses about how systems develop.

This work responds to that challenge by uniting recent advances from systems engineering, complex systems, engineering design, and organization science to form the ComplEx System Integrated Utilities Model. Beginning from a review of terminology, the piece identifies the abstracted elements composing development processes and integrates techniques for simulating each aspect into a modeling framework. It then demonstrates the ability of CESIUM to generate sizeable datasets representing the outcomes of system development processes from which researchers can form theory. The underlying flexibility of the framework leaves numerous opportunities for researchers to explore abstracted phenomena and real-world examples alike toward the construction of systems engineering theory.

\section*{Acknowledgment}
Thanks to all of our colleagues and friends who reviewed this work and supported us, including Abigail Jacobs, Vanessa Rojano, Mojtaba Arezoomand, Lynette Shaw, and Maciej Kurant.

\bibliographystyle{IEEEtran}

\bibliography{ms}


\end{document}

%% file: figures/figure_network.tex
\begin{figure}
    \centering
    
        \subfloat[]{\includegraphics[width=0.38\textwidth]{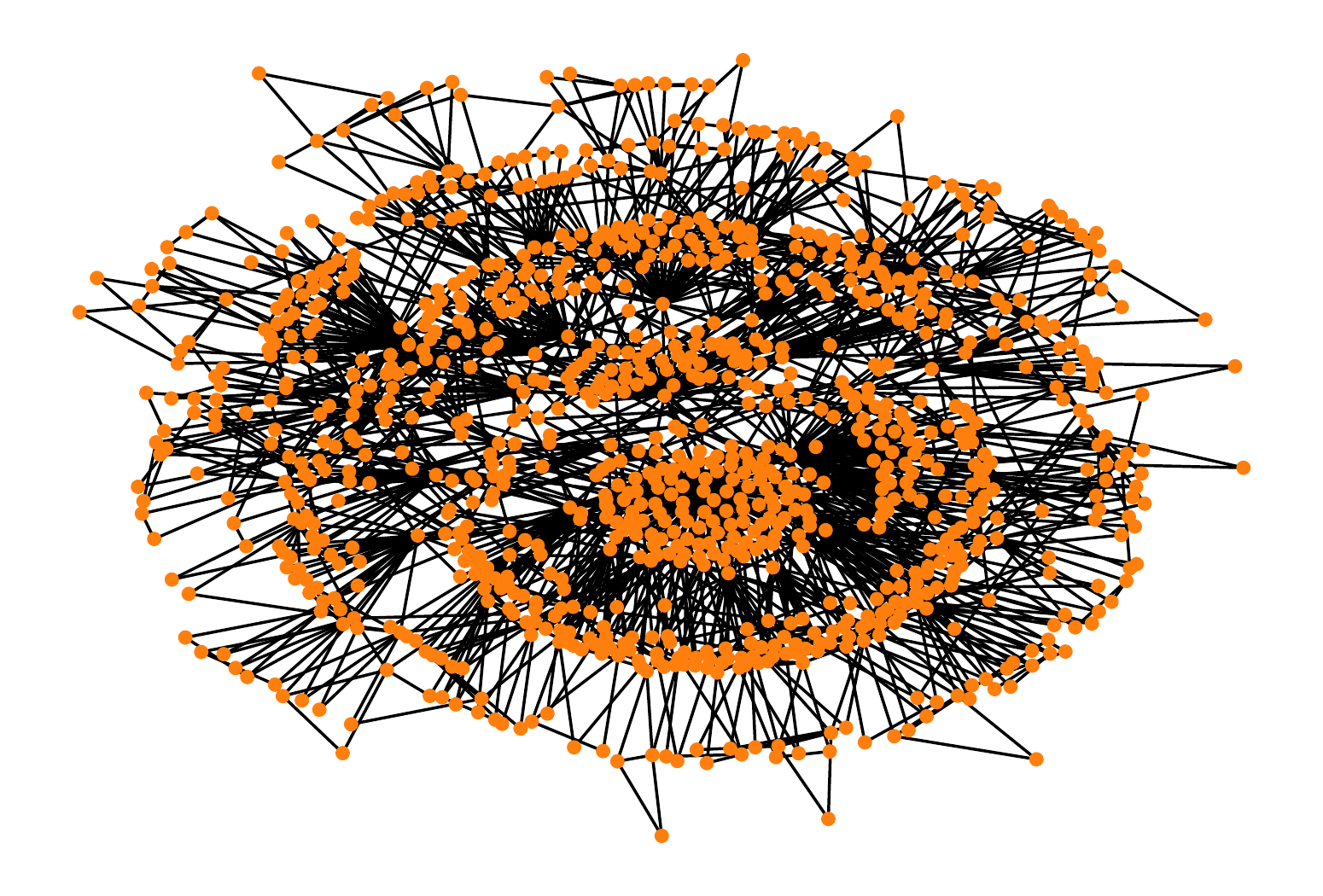}
        \label{fig:network_graph}}
        
        \subfloat[]{\includegraphics[width=0.38\textwidth]{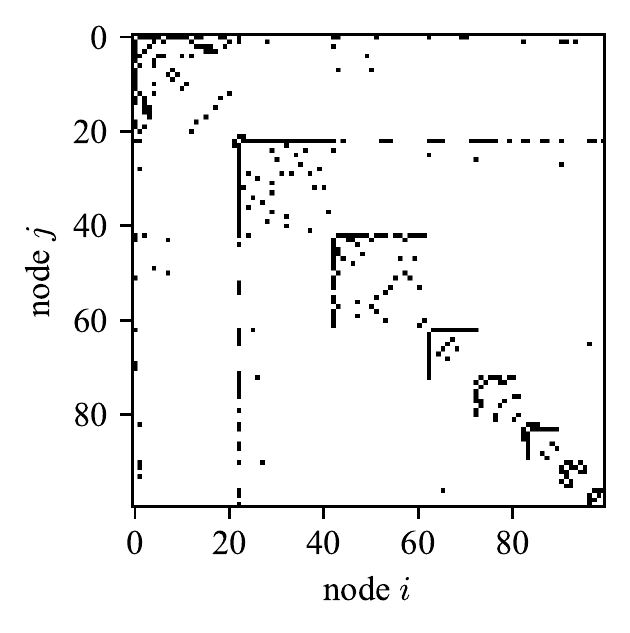}
        \label{fig:adjacency_matrix}}
        
        \subfloat[]{\includegraphics[width=0.38\textwidth]{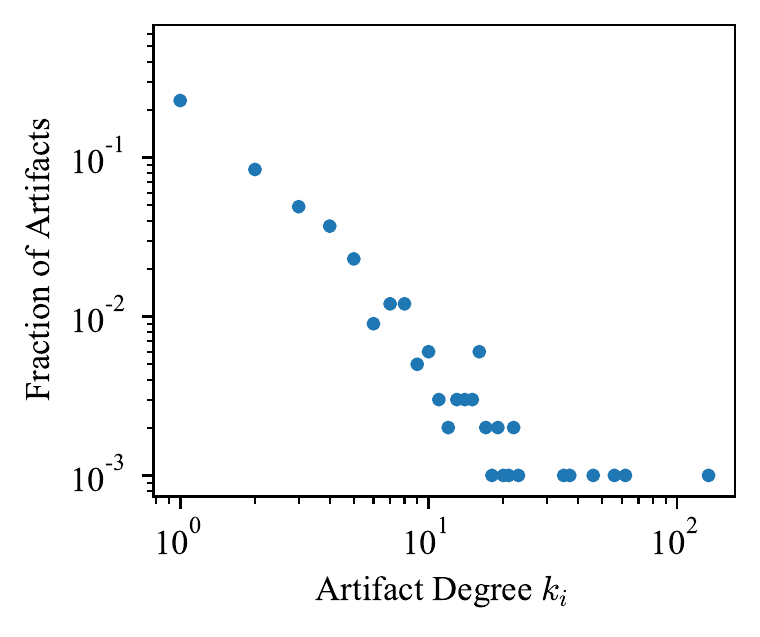}
        \label{fig:degree_distribution}}
        
    \caption{Visual representations of undirected networks generated with a Holme-Kim preferential attachment algorithm and probability that new edges will form a triangle $p_t=0.9$. (a) Graph of a network with $n=1000$ artifacts. The orange dots represent nodes or artifacts, and the black lines represent edges or interfaces. (b) An adjacency matrix $A_{ij}$ for a network with $n=100$ artifacts where $A_{ij}=1$ (black) if an edge exists between nodes $i$ and $j$ and $0$ (white) if not. Also known as a Design Structure Matrix. (c) Scale-free degree distribution of a network with $n=1000$ artifacts. Note the approximately-linear, negatively-sloping form of the distribution on a log-log scale, characteristic of a scale-free degree distribution \cite{Newman2018}.}
    \label{fig:verification_construction}
\end{figure}

%% file: figures/figure_interaction.tex
\begin{figure}
    \centering
        \includegraphics[width=0.48\textwidth]{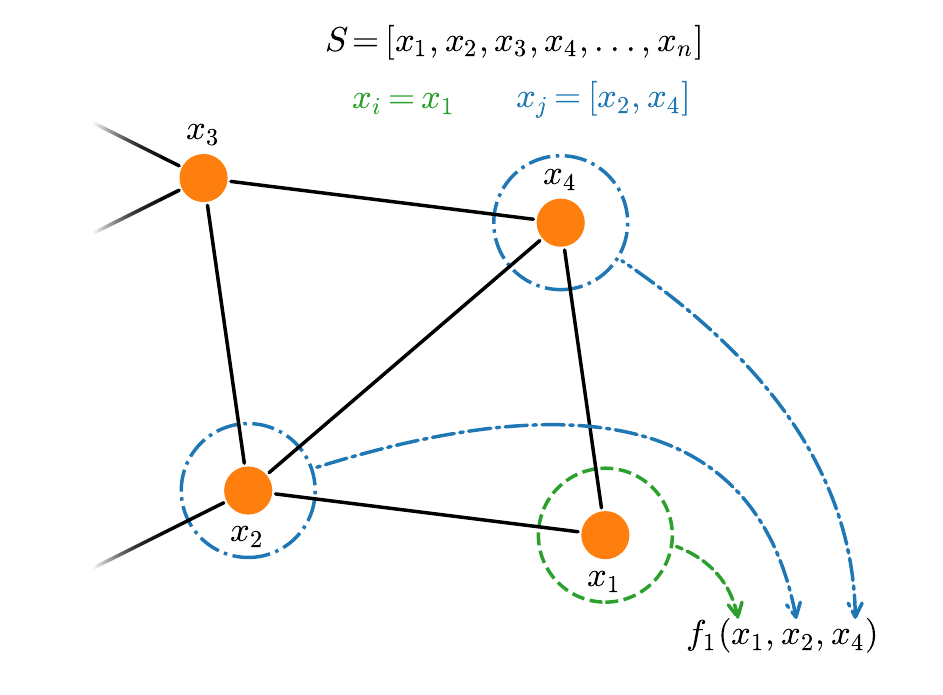}
        \caption{An example of artifact interaction. In this case, the $i^{th}$ artifact is artifact 1 with variable $x_i=x_1$. Artifact 1 interacts with $j\in \{2,4\}$ and so $x_j=[x_2,x_4]$. Therefore, artifact 1's objective function is $f_1(x_1,x_2,x_4)$.}
        \label{fig:interaction}
\end{figure}

%% file: figures/figure_objectives.tex
\begin{figure*}
    \centering
        \subfloat[]{\includegraphics[width=0.24\textwidth]{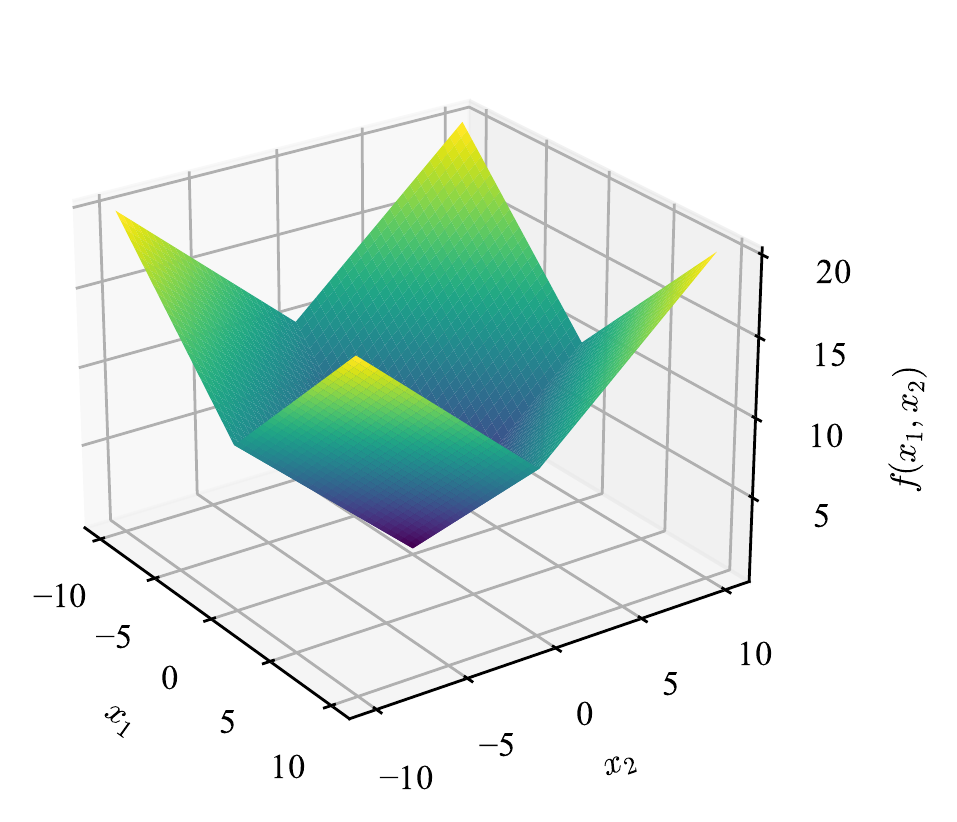}
        \label{fig:fn_abssum}}
        \subfloat[]{\includegraphics[width=0.24\textwidth]{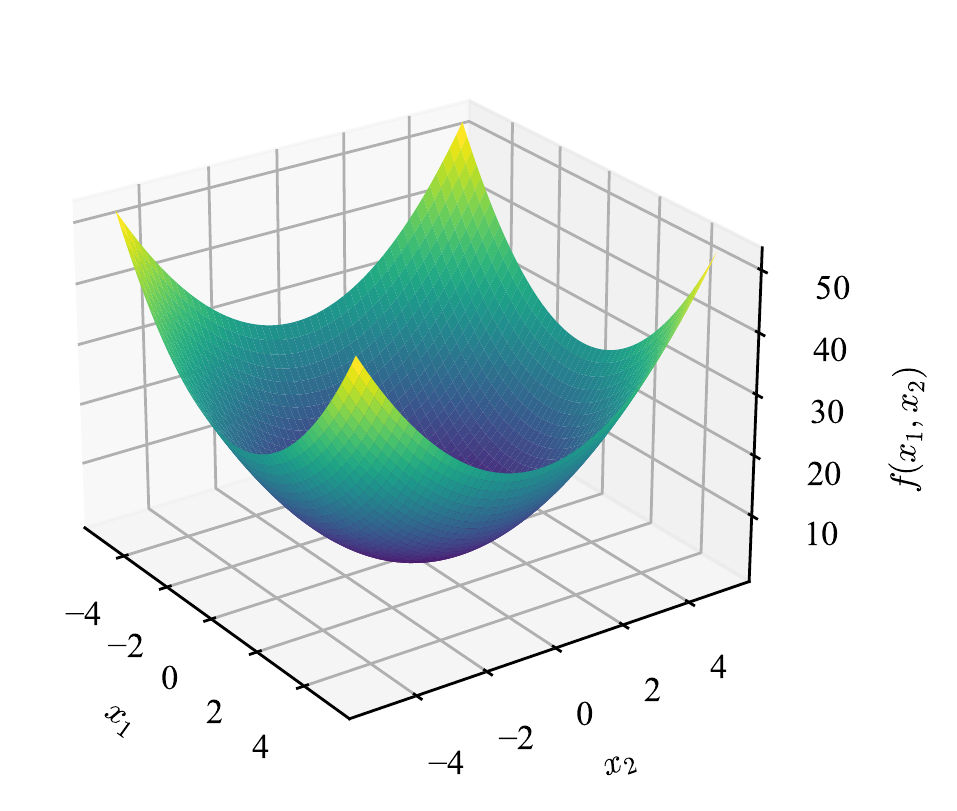}
        \label{fig:fn_sphere}}
        \subfloat[]{\includegraphics[width=0.24\textwidth]{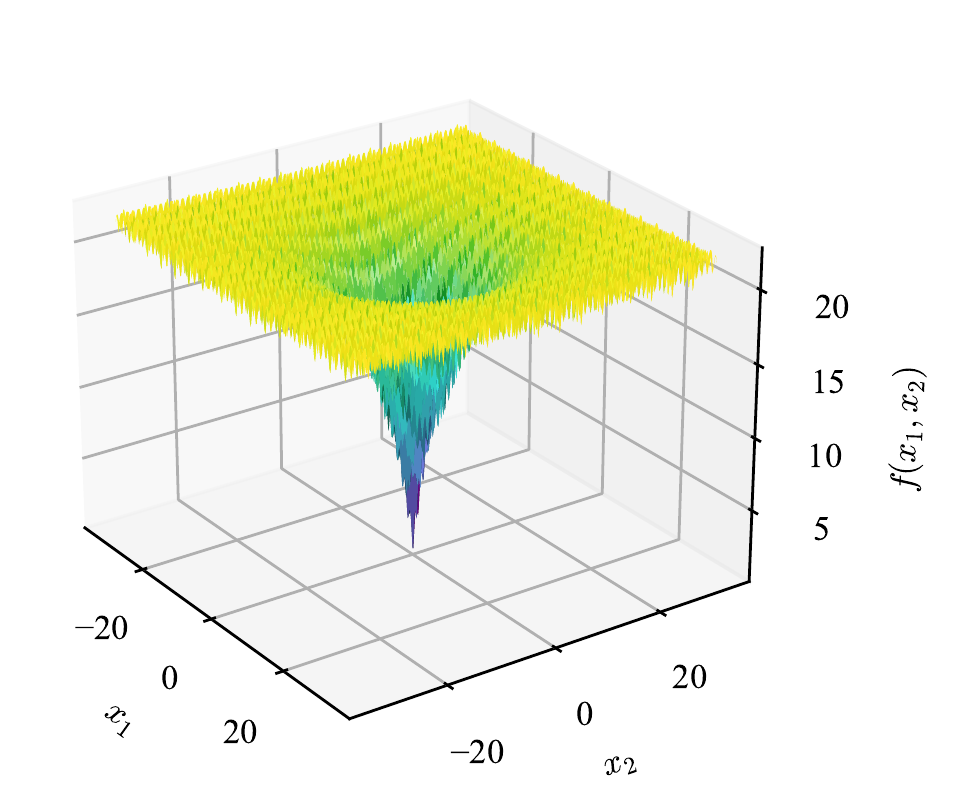}
        \label{fig:fn_ackley}}
        \subfloat[]{\includegraphics[width=0.24\textwidth]{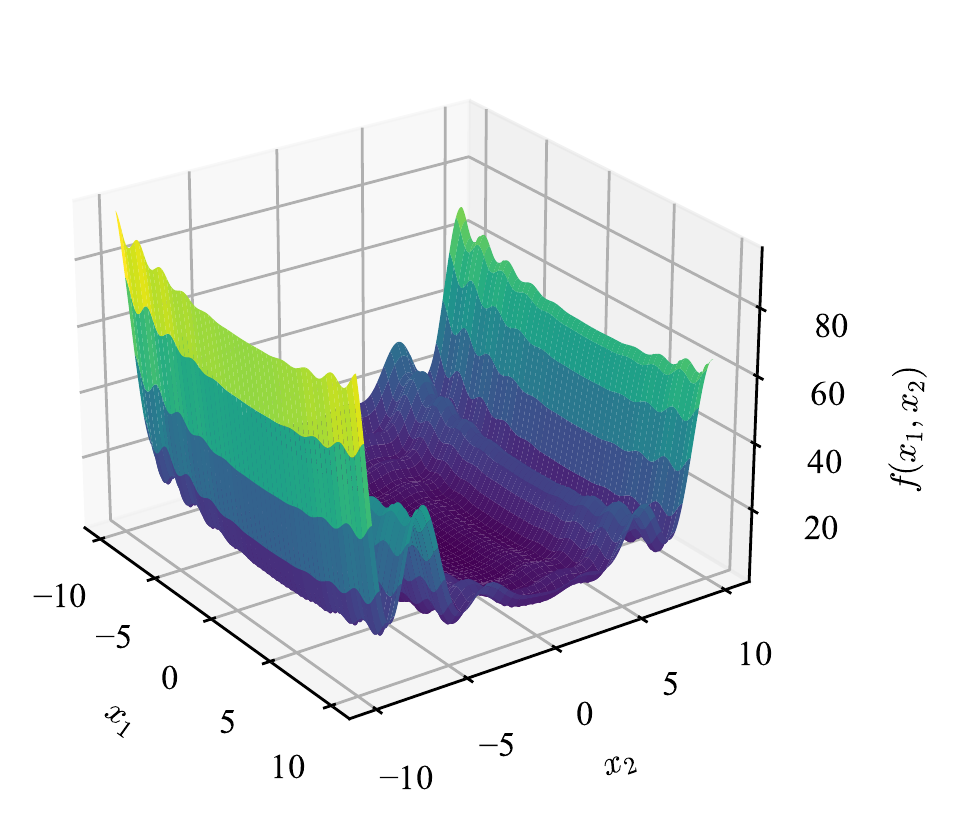}
        \label{fig:fn_levy}}
    \caption{Graphical representations of the selected objective functions with two decision variables. (a) Absolute Sum Function. (b) Sphere Function. (c) Ackley Function. (d) Levy Function. }
    \label{fig:objective_functions}
\end{figure*}

%% file: figures/eq_absolute-sum.tex
The Absolute Sum function, corresponding to the first term of the Taylor expansion, on domain for all $x_m\in [-10.00,10.00]$:
\begin{equation}
    f_i(\Vec{x}_i)= \sum_{m=1}^{k_i+1} |x_m|
\end{equation}
The optimum $\Vec{x}_i^*=(0,\ldots,0)$ minimizes $f_i$ for the Absolute Sum function yielding $f_i(\Vec{x}_i^*)=0$.

%% file: figures/eq_sphere.tex
The Sphere function \cite{Surjanovic2013}, corresponding to the second-order term of the Taylor expansion, on the recommended evaluation domain for all $x_m\in [-5.12,5.12]$:
\begin{equation}
    f_i(\Vec{x}_i)=\sum_{m=1}^{k_i+1} x_m^2
\end{equation}
with $f_i(\Vec{x}_i^*)=0$ at $\Vec{x}_i^*=(0,\ldots,0)$.

%% file: figures/eq_ackley.tex
The Ackley function \cite{Surjanovic2013}, a symmetrical function with multiple minima, on the recommended evaluation domain for all $x_m\in [-32.768,32.768]$, $c_1=20$, $c_2=0.2$, and $c_3=2\pi$:
\begin{equation}
    \begin{split}
        f_i(\Vec{x}_i)=&-c_1\exp\Bigg(-c_2\sqrt{\frac{1}{k_i+1}\sum_{m=1}^{k_i+1} x_m^2}\Bigg) \\
        & -\exp\Bigg(\frac{1}{k_i+1}\sum_{m=1}^{k_i+1} \cos(c_{3}x_m)\Bigg)+e^1+c_1
    \end{split}
\end{equation}
with $f_i(\Vec{x}_i^*)=0$ at $\Vec{x}_i^*=(0,\ldots,0)$.

%% file: figures/eq_levy.tex
The Levy function \cite{Surjanovic2013}, an asymmetrical function with multiple minima, on the recommended evaluation domain for all $x_m\in [-10.00,10.00]$, and $c_m=1+\frac{x_m-1}{4}$:
\begin{equation}
    \begin{split}
        f_i\left(\Vec{x}_i\right)=&\sin^2\left(\pi c_1\right) \\
        & + \sum_{m=1}^{k_i} \left(c_m-1\right)^2\left(1+10\sin^2\left(\pi c_m + 1\right)\right) \\
        & + \left(c_{k_i+1}-1\right)^2\left(1+\sin^2\left(2\pi c_{k_i+1}\right)\right)
    \end{split}
\end{equation}
with $f_i(\Vec{x}_i^*)=0$ at $\Vec{x}_i^*=(1,\ldots,1)$, though a variable transformation could shift this optima to the origin.

%% file: figures/figure_example.tex
\begin{figure*}
    \centering
    \subfloat[]{\includegraphics[width=0.32\textwidth]{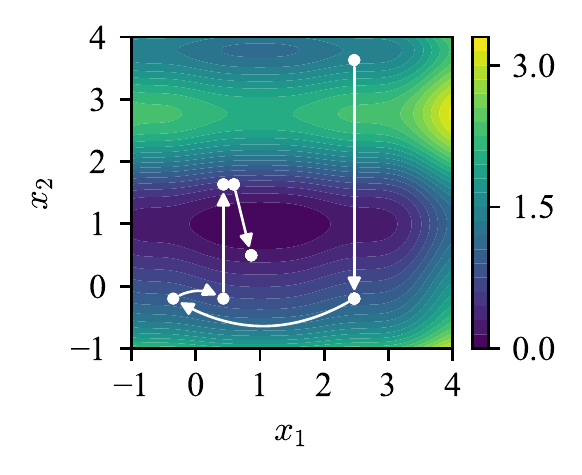}
    \label{fig:example_f1}}
    \subfloat[]{\includegraphics[width=0.32\textwidth]{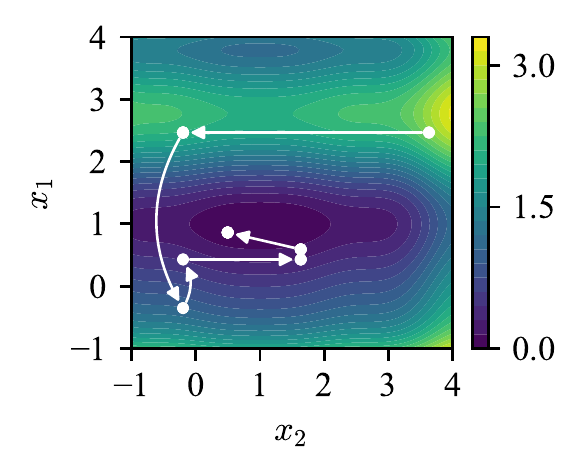}
    \label{fig:example_f2}}
    \subfloat[]{\includegraphics[width=0.32\textwidth]{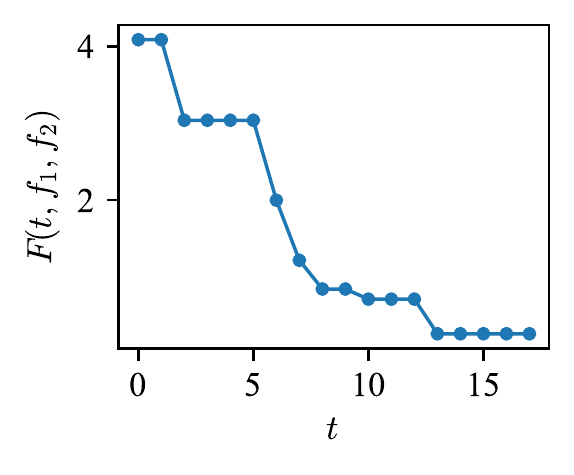}
    \label{fig:example_system}}
    \caption{An example execution of CESIUM for a system with two agents using the Levy function. White dots and arrows illustrate the progress of $\Vec{x}$ over time, thereby affecting the objective evaluations (a) $f_1$ of agent 1, (b) $f_2$ of agent 2, and (c) $F$ the overall system performance. (a) $f_1(x_1,x_2)$ (b) $f_2(x_2,x_1)$ (c) $F(t,f_1,f_2)$}
    \label{fig:example_execution}
\end{figure*}

%% file: figures/table_parameters.tex
\begin{table*}
\begin{singlespace}
\begin{footnotesize}
\centering 
\caption{Independent and dependent variables for the base instance of CESIUM.} 
\label{tab:parameters} 
\renewcommand{\arraystretch}{1.5}
\begin{tabular}{p{0.13\linewidth} p{0.05\linewidth} p{0.1\linewidth} p{0.08\linewidth} p{0.33\linewidth} p{0.15\linewidth}} 
\toprule
Name
& Variable
& Abbreviation
& Type 
& Description & Values \\
\midrule

\makecell[tl]{Num. of Nodes}
& \makecell[tc]{$n$}
& \makecell[tl]{Num. Nodes}
& Independent
& The number of nodes in the generated system during that execution.
& $\{$50, 100, 500, 1000$\}$
\\

\makecell[tl]{Objective Function}
& \makecell[tc]{$f_i$}
& \makecell[tl]{Obj. Fn.}
& Independent
& The objective function employed by each agent to evaluate their individual performance.
& \makecell[tl]{$\{$absolute-sum, sphere, \\ levy, ackley$\}$}
\\

\makecell[tl]{Triangle Probability}
& \makecell[tc]{$p_t$}
& \makecell[tl]{Tri. Prob.}
& Independent
& An input to the Holme-Kim preferential attachment algorithm along with the Num. of Edges Added. The probability that a new edge originating from node $i$ will be placed such that the new edge connects $i$ to a node $l$ which is already connected to node $j$, thereby forming a triangle between $i$, $j$, and $l$. The reverse, $1-$x\_prob\_triangle, is the probability that new nodes will connect to \textit{any} existing node according to preferential attachment.
& $\{$0, 0.1, 0.2, \ldots, 1.0$\}$
\\

\makecell[tl]{Convergence \\ Threshold}
& \makecell[tc]{$\varepsilon$}
& \makecell[tl]{Conv. Thresh.}
& Independent
& The system convergence threshold.
& \makecell[tl]{$\{$0.01, 0.05, 0.1, 0.5, \\ 1.0, 5.0, 10.0$\}$}
\\

\makecell[tl]{Future Estimate \\ Probability}
& \makecell[tc]{$p_e$}
& \makecell[tl]{Fut. Est. \\ Prob.}
& Independent
& Probability that a given agent will utilize a ``future'' estimate (when $\mu$ is ``future'') throughout the system execution, where $1-p_e$ is the probability that an agent will utilize a ``current'' estimate instead.
& $\{$0, 0.1, 0.2, \ldots, 1.0$\}$
\\

\makecell[tl]{Num. of Edges \\ Added}
& \makecell[tc]{$h$}
& \makecell[tl]{Num. Edg.}
& Constant
& The number of edges generated to connect new nodes $i$ to existing nodes $j$ in the system during network generation using the Holme-Kim preferential attachment algorithm.
& 2
\\

\makecell[tl]{Maximum Num. \\ of Design Cycles}
& \makecell[tc]{$d$}
& \makecell[tl]{Max. Cycl.}
& Constant
& The maximum number of design cycles.
& 100
\\

\makecell[tl]{Initial Temperature}
& \makecell[tc]{$\tau$}
& \makecell[tl]{Init. Temp.}
& Constant
& The initial temperature for the dual annealing algorithm in SciPy \cite{Virtanen2020}.
& 0.1
\\

\makecell[tl]{Num. of Annealing \\ Iterations}
& \makecell[tc]{$\omega$}
& \makecell[tl]{Anneal. Iter.}
& Constant
& The maximum number of global search iterations for the dual annealing algorithm.
& 1
\\

\makecell[tl]{Estimation Method}
& \makecell[tc]{$\mu$}
& \makecell[tl]{Est. Meth.}
& Constant
& Specifies the types of estimates generated by agents. A value of ``current'' allows only current design values. A value of ``future'' allows probabilistic selection of either ``current'' design values of $x_i$ or ``future'' predictions of $x_i$.
& ``future''
\\

\makecell[tl]{Cooling Rate}
& \makecell[tc]{$\rho$}
& \makecell[tl]{Cool. Rate}
& Constant
& Cooling rate of the dual annealing algorithm.
& 2.62
\\

\makecell[tl]{Num. of Design \\ Cycles}
& \makecell[tc]{$N$}
& \makecell[tl]{Num. Cycl.}
& Dependent
& The number of design cycles that it took the specified execution of CESIUM to converge, bounded by the minimum number of cycles required for convergence and the maximum allowed design cycles.
& [4,100]
\\

\makecell[tl]{System Performance}
& \makecell[tc]{$F$}
& \makecell[tl]{Sys. Perf.}
& Dependent
& The final system performance for the specified execution.
& \makecell[tl]{Distribution- and \\ objective-specific}
\\

\bottomrule
\end{tabular} 
\end{footnotesize}
\end{singlespace}
\end{table*}

%% file: figures/table_reg_perf.tex
\begin{table*}
\begin{singlespace}
\begin{footnotesize}
\centering 
\caption{Multivariate OLS regressions with robust standard errors of the varied parameters against System Performance. Models II.1 \& II.2 include objective function fixed effects; Models II.3 through II.7 instead analyze the objective function subsets of the dataset. More negative values represent more optimal performances of objective functions.} 
\label{tab:regression_performance}
\begin{center}
\begin{tabular}[t]{lccccccc}
\toprule
\multicolumn{1}{c}{ } & \multicolumn{2}{c}{All Functions} & \multicolumn{1}{c}{Absolute Sum} & \multicolumn{2}{c}{Sphere} & \multicolumn{1}{c}{Ackley} & \multicolumn{1}{c}{Levy} \\
\cmidrule(l{3pt}r{3pt}){2-3} \cmidrule(l{3pt}r{3pt}){4-4} \cmidrule(l{3pt}r{3pt}){5-6} \cmidrule(l{3pt}r{3pt}){7-7} \cmidrule(l{3pt}r{3pt}){8-8}
  & Model II.1 & Model II.2 & Model II.3 & Model II.4 & Model II.7 & Model II.5 & Model II.6\\
\midrule
Dependent Variable & Sys. Perf. & Sys. Perf. & Sys. Perf. & Sys. Perf. & log(Sys. Perf.) & Sys. Perf. & Sys. Perf.\\
\midrule
Constant & -1563.787*** & -226.174*** & 0.450*** & 13.298*** & 1.972*** & -3.004*** & 615.742***\\
 & (6.696) & (3.125) & (0.041) & (0.499) & (0.032) & (0.568) & (10.110)\\
Fn: Sphere & 9.355+ & -30.709*** &  &  &  &  & \\
 & (3.997) & (1.090) &  &  &  &  & \\
Fn: Ackley & 303.090*** & 20.903*** &  &  &  &  & \\
 & (3.714) & (0.992) &  &  &  &  & \\
Fn: Levy & 7477.843*** & 1540.989*** &  &  &  &  & \\
 & (8.969) & (6.987) &  &  &  &  & \\
Num. Nodes\textsuperscript{$\dagger$} & 4.301*** & 0.529*** & 0.000 & -0.042*** & -0.321*** & 0.731*** & 18.636***\\
 & (0.008) & (0.005) & (0.000) & (0.001) & (0.006) & (0.001) & (0.022)\\
Tri. Prob. & 6.884 & -17.531*** & -0.250*** & -27.972*** & -1.371*** & -2.032+ & -12.871\\
 & (7.675) & (4.758) & (0.070) & (0.838) & (0.047) & (0.956) & (16.545)\\
Conv. Thresh.\textsuperscript{$\dagger$} & 17.287*** & 21.547*** & 3.011*** & 1.697*** & 1.219*** & 15.859*** & 123.995***\\
 & (0.679) & (0.533) & (0.015) & (0.101) & (0.007) & (0.148) & (1.930)\\
Fut. Est. Prob. & -496.031*** & 398.783*** & -0.278*** & -25.708*** & -3.399*** & -1.357 & -362.163***\\
 & (7.735) & (5.172) & (0.070) & (0.944) & (0.048) & (0.968) & (16.576)\\
Num. Nodes\textsuperscript{$\dagger$} $\times$ Tri. Prob. &  & 0.013 & 0.000 & 0.033*** & 0.201*** & -0.023*** & 0.041\\
 &  & (0.007) & (0.000) & (0.001) & (0.008) & (0.002) & (0.027)\\
Num. Nodes\textsuperscript{$\dagger$} $\times$ Conv. Thresh.\textsuperscript{$\dagger$} &  & -0.031*** & -0.001*** & 0.000*** & -0.029*** & -0.019*** & -0.104***\\
 &  & (0.001) & (0.000) & (0.000) & (0.001) & (0.000) & (0.002)\\
Num. Nodes\textsuperscript{$\dagger$} $\times$ Fut. Est. Prob. &  & -0.927*** & 0.000** & 0.107*** & 0.688*** & -0.014*** & -3.800***\\
 &  & (0.007) & (0.000) & (0.001) & (0.009) & (0.002) & (0.027)\\
Tri. Prob. $\times$ Conv. Thresh.\textsuperscript{$\dagger$} &  & 0.021 & 0.028 & 0.048 & -0.055*** & 0.326 & -0.318\\
 &  & (0.637) & (0.017) & (0.103) & (0.004) & (0.179) & (2.460)\\
Tri. Prob. $\times$ Fut. Est. Prob. &  & 24.695*** & 0.502*** & 57.744*** & 1.070*** & 4.366* & 36.168\\
 &  & (7.133) & (0.118) & (1.547) & (0.032) & (1.616) & (26.805)\\
Conv. Thresh.\textsuperscript{$\dagger$} $\times$ Fut. Est. Prob. &  & -12.102*** & 0.016 & -0.005 & -0.161*** & 0.002 & -48.423***\\
 &  & (0.651) & (0.018) & (0.134) & (0.005) & (0.181) & (2.462)\\
Fn: Sphere $\times$ Num. Nodes &  & 0.029*** &  &  &  &  & \\
 &  & (0.001) &  &  &  &  \vphantom{1} & \\
Fn: Sphere $\times$ Tri. Prob. &  & 14.602*** &  &  &  &  & \\
 &  & (1.037) &  &  &  &  & \\
Fn: Sphere $\times$ Conv. Thresh. &  & -1.170*** &  &  &  &  & \\
 &  & (0.103) &  &  &  &  & \\
Fn: Sphere $\times$ Fut. Est. Prob. &  & 47.295*** &  &  &  &  & \\
 &  & (1.354) &  &  &  &  & \\
Fn: Ackley $\times$ Num. Nodes &  & 0.669*** &  &  &  &  & \\
 &  & (0.001) &  &  &  &  & \\
Fn: Ackley $\times$ Tri. Prob. &  & -8.667*** &  &  &  &  & \\
 &  & (1.027) &  &  &  &  & \\
Fn: Ackley $\times$ Conv. Thresh. &  & 5.457*** &  &  &  &  & \\
 &  & (0.104) &  &  &  &  & \\
Fn: Ackley $\times$ Fut. Est. Prob. &  & -5.120*** &  &  &  &  & \\
 &  & (1.276) &  &  &  &  & \\
Fn: Levy $\times$ Num. Nodes &  & 16.512*** &  &  &  &  & \\
 &  & (0.009) &  &  &  &  & \\
Fn: Levy $\times$ Tri. Prob. &  & 21.064+ &  &  &  &  & \\
 &  & (8.670) &  &  &  &  & \\
Fn: Levy $\times$ Conv. Thresh. &  & 54.089*** &  &  &  &  & \\
 &  & (0.793) &  &  &  &  & \\
Fn: Levy $\times$ Fut. Est. Prob. &  & -2026.815*** &  &  &  &  & \\
 &  & (8.833) &  &  &  &  & \\
\midrule
Num. Obs. & 1355200 & 1355200 & 338800 & 338800 & 338800 & 338800 & 338800\\
R-squared & 0.618 & 0.969 & 0.722 & 0.125 & 0.663 & 0.886 & 0.945\\
R-squared Adj. & 0.618 & 0.969 & 0.722 & 0.125 & 0.663 & 0.886 & 0.945\\
Residual Standard Error & 2824.141 & 798.934 & 5.935 & 64.654 & 1.617 & 92.402 & 1537.809\\
\bottomrule
\multicolumn{8}{l}{Significances: + p $<$ 0.05, * p $<$ 0.01, ** p $<$ 0.005, *** p $<$ 0.001}\\
\multicolumn{8}{l}{Robust standard errors in parentheses.}\\
\multicolumn{8}{l}{\textsuperscript{$\dagger$} Variable log-transformed for Model II.7 only}\\
\end{tabular}
\end{center}
\end{footnotesize}
\end{singlespace}
\end{table*}

%% file: figures/table_reg_cycl.tex
\begin{table*}
\begin{singlespace}
\begin{footnotesize}
\centering 
\caption{Multivariate OLS regressions with robust standard errors of the varied parameters against the Number of Design Cycles. Models III.1 \& III.2 include objective function fixed effects; Models III.3 through III.6 instead analyze the objective function subsets of the dataset.} 
\label{tab:regression_cycles}
\begin{center}
\begin{tabular}[t]{lcccccc}
\toprule
\multicolumn{1}{c}{ } & \multicolumn{2}{c}{All Functions} & \multicolumn{1}{c}{Absolute Sum} & \multicolumn{1}{c}{Sphere} & \multicolumn{1}{c}{Ackley} & \multicolumn{1}{c}{Levy} \\
\cmidrule(l{3pt}r{3pt}){2-3} \cmidrule(l{3pt}r{3pt}){4-4} \cmidrule(l{3pt}r{3pt}){5-5} \cmidrule(l{3pt}r{3pt}){6-6} \cmidrule(l{3pt}r{3pt}){7-7}
  & Model III.1 & Model III.2 & Model III.3 & Model III.4 & Model III.5 & Model III.6\\
\midrule
Dependent Variable & Num. Cycles & Num. Cycles & Num. Cycles & Num. Cycles & Num. Cycles & Num. Cycles\\
\midrule
Constant & 63.659*** & 66.030*** & 67.403*** & 42.565*** & 98.085*** & 45.007***\\
 & (0.038) & (0.064) & (0.100) & (0.060) & (0.051) & (0.197)\\
Fn: Sphere & -22.043*** & -21.819*** &  &  &  & \\
 & (0.023) & (0.062) &  &  &  & \\
Fn: Ackley & 27.487*** & 30.553*** &  &  &  & \\
 & (0.027) & (0.062) &  &  &  & \\
Fn: Levy & -6.666*** & -21.809*** &  &  &  & \\
 & (0.074) & (0.176) &  &  &  & \\
Num. Nodes & 0.021*** & 0.025*** & 0.027*** & 0.019*** & 0.010*** & 0.134***\\
 & (0.000) & (0.000) & (0.000) & (0.000) & (0.000) & (0.001)\\
Tri. Prob. & -0.130** & 0.112 & -0.290+ & -0.057 & -0.034 & -1.588***\\
 & (0.040) & (0.087) & (0.122) & (0.090) & (0.078) & (0.274)\\
Conv. Thresh. & -2.648*** & -3.977*** & -3.004*** & -2.219*** & -2.905*** & -5.185***\\
 & (0.004) & (0.027) & (0.096) & (0.008) & (0.014) & (0.026)\\
Fut. Est. Prob. & 1.894*** & -0.656*** & -0.199 & -0.044 & -0.107 & 8.884***\\
 & (0.040) & (0.087) & (0.130) & (0.090) & (0.078) & (0.283)\\
Num. Nodes $\times$ Tri. Prob. &  & 0.000*** & 0.000*** & 0.000 & -0.003*** & -0.011***\\
 &  & (0.000) & (0.000) & (0.000) & (0.000) & \vphantom{1} (0.001)\\
Num. Nodes $\times$ Conv. Thresh. &  & 0.001*** & -0.001*** & 0.000*** & 0.008*** & 0.005***\\
 &  & (0.000) & (0.000) & (0.000) & (0.000) & (0.000)\\
Num. Nodes $\times$ Fut. Est. Prob. &  & 0.000+ & 0.000** & 0.000*** & 0.000 & -0.020***\\
 &  & (0.000) & (0.000) & (0.000) & (0.000) & (0.001)\\
Tri. Prob. $\times$ Conv. Thresh. &  & -0.002 & -0.509*** & -0.015 & -0.106*** & -0.010\\
 &  & (0.009) & (0.105) & (0.009) & (0.015) & (0.028)\\
Tri. Prob. $\times$ Fut. Est. Prob. &  & 0.068 & 0.165 & 0.006 & 0.084 & 1.234***\\
 &  & (0.104) & (0.153) & (0.124) & (0.115) & (0.372)\\
Conv. Thresh. $\times$ Fut. Est. Prob. &  & 0.307*** & -0.503*** & -0.060*** & -0.046** & 0.736***\\
 &  & (0.009) & (0.150) & (0.009) & (0.015) & (0.032)\\
Fn: Sphere $\times$ Num. Nodes &  & -0.008*** &  &  &  & \\
 &  & (0.000) &  &  &  \vphantom{2} & \\
Fn: Sphere $\times$ Tri. Prob. &  & -0.007 &  &  &  & \\
 &  & (0.069) &  &  &  \vphantom{1} & \\
Fn: Sphere $\times$ Conv. Thresh. &  & 0.943*** &  &  &  & \\
 &  & (0.026) &  &  &  \vphantom{1} & \\
Fn: Sphere $\times$ Fut. Est. Prob. &  & 0.004 &  &  &  & \\
 &  & (0.069) &  &  &  & \\
Fn: Ackley $\times$ Num. Nodes &  & 0.012*** &  &  &  & \\
 &  & (0.000) &  &  &  \vphantom{1} & \\
Fn: Ackley $\times$ Tri. Prob. &  & -0.415*** &  &  &  & \\
 &  & (0.071) &  &  &  \vphantom{1} & \\
Fn: Ackley $\times$ Conv. Thresh. &  & 1.559*** &  &  &  & \\
 &  & (0.026) &  &  &  & \\
Fn: Ackley $\times$ Fut. Est. Prob. &  & 0.112 &  &  &  & \\
 &  & (0.071) &  &  &  & \\
Fn: Levy $\times$ Num. Nodes &  & 0.104*** &  &  &  & \\
 &  & (0.000) &  &  &  & \\
Fn: Levy $\times$ Tri. Prob. &  & -1.022*** &  &  &  & \\
 &  & (0.204) &  &  &  & \\
Fn: Levy $\times$ Conv. Thresh. &  & -1.104*** &  &  &  & \\
 &  & (0.031) &  &  &  & \\
Fn: Levy $\times$ Fut. Est. Prob. &  & 11.345*** &  &  &  & \\
 &  & (0.207) &  &  &  & \\
\midrule
Num. Obs. & 1355200 & 1355200 & 338800 & 338800 & 338800 & 338800\\
R-squared & 0.724 & 0.813 & 0.795 & 0.709 & 0.726 & 0.677\\
R-squared Adj. & 0.724 & 0.813 & 0.795 & 0.709 & 0.726 & 0.677\\
Residual Standard Error & 14.628 & 12.046 & 8.875 & 7.041 & 6.699 & 18.834\\
\bottomrule
\multicolumn{7}{l}{Significances: + p $<$ 0.05, * p $<$ 0.01, ** p $<$ 0.005, *** p $<$ 0.001}\\
\multicolumn{7}{l}{Robust standard errors in parentheses.}\\
\end{tabular} 
\end{center}
\end{footnotesize}
\end{singlespace}
\end{table*}

%% file: figures/figure_feat_importances.tex
\begin{figure*}
    \centering
        \includegraphics{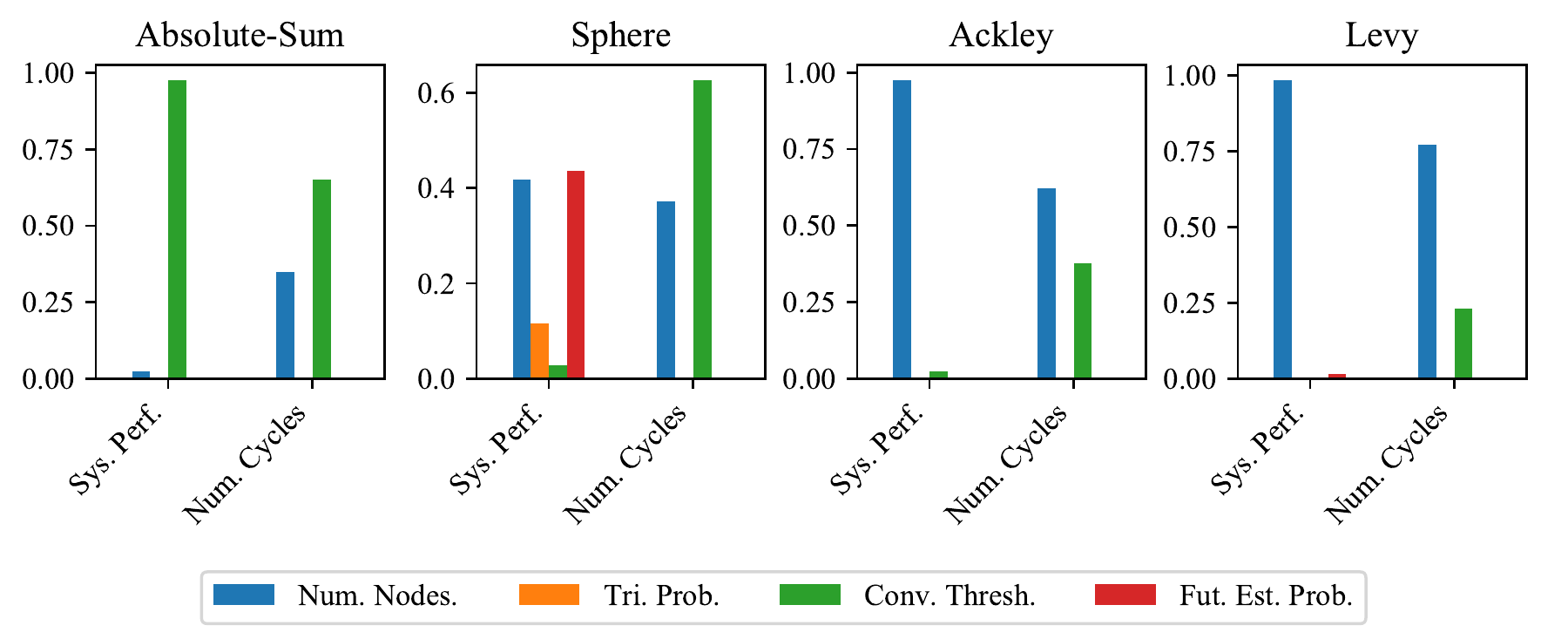}
        \caption{Random forest feature importances for the objective function subsets of the executions. Higher bars indicate a more important feature (greater reduction in error when predicting the outcome). The Number of Nodes and the Convergence Threshold tended to have the greatest importance on System Performance and the Number of Design Cycles for most objective functions, the notable exception being the Sphere Function, the outcomes of which were significantly affected by the Future Estimate Probability as well.}
        \label{fig:feat_importances}
\end{figure*}